\documentclass[11pt,notitlepage]{article}

\usepackage[utf8]{inputenc}
\usepackage{amsmath}
\usepackage{amsfonts}
\usepackage{graphics}
\usepackage{graphicx}
\usepackage{amssymb}
\usepackage{epigraph}
\usepackage{latexsym}
\usepackage{ifthen}
\usepackage{setspace}
\usepackage{epsfig}
\usepackage{multirow}
\usepackage{booktabs}
\usepackage{array}
\usepackage{float}
\usepackage{verbatim}
\usepackage{rotating}
\usepackage{caption}
\usepackage{subcaption}
\usepackage{comment}
\usepackage{stata}
\usepackage{bm}
\usepackage{tikz}
\usetikzlibrary{arrows,decorations.pathmorphing,backgrounds,positioning,fit,petri,matrix,trees,shadows}
\usepackage{authblk}
\usepackage{natbib}
\usepackage{geometry}
\title{A flexible parametric accelerated failure time model}
\author[1,2,$\star$]{Michael J. Crowther}
\author[3]{Patrick Royston}
\author[2]{Mark Clements}
\affil[1]{University of Leicester, Biostatistics Research Group, Department of Health Sciences, University Road, Leicester, LE1 7RH, UK.}
\affil[2]{Karolinska Institutet, Department of Medical Epidemiology and Biostatistics, Box 281, S-171 77 Stockholm, Sweden.}
\affil[3]{MRC CTU at UCL, 90 High Holborn, Holborn, London WC1V 6LJ, UK.}
\affil[$\star$]{michael.crowther@le.ac.uk}
\date{}

\begin{document}

\maketitle

\begin{abstract}
Accelerated failure time (AFT) models are used widely in medical research, though to a much lesser extent than proportional hazards models. In an AFT model, the effect of covariates act to accelerate or decelerate the time to event of interest, i.e. shorten or extend the time to event. Commonly used parametric AFT models are limited in the underlying shapes that they can capture. In this article, we propose a general parametric AFT model, and in particular concentrate on using restricted cubic splines to model the baseline to provide substantial flexibility. We then extend the model to accommodate time-dependent acceleration factors. Delayed entry is also allowed, and hence, time-dependent covariates. We evaluate the proposed model through simulation, showing substantial improvements compared to standard parametric AFT models. We also show analytically and through simulations that the AFT models are collapsible, suggesting that this model class will be well suited to causal inference. We illustrate the methods with a dataset of patients with breast cancer. User friendly Stata and R software packages are provided.
\end{abstract}

\section{Introduction}
\label{sec:intro}

Accelerated failure time (AFT) models are commonly used in a variety of settings within the medical literature \citep{Collettbook2003}. The interpretation of an acceleration factor can be considered more intuitive, directly adjusting the survival time, either increasing or decreasing it, compared to the interpretation of a hazard ratio, meaning a relative increase or decrease in the event rate \citep{Swindell2009}. A parametric approach tends to be favoured when fitting an AFT model; however, parametric models are limited by the flexibility of the distribution chosen \citep{Cox2007,Cox2008}. Parametric AFT models are particular prevalent in economic decision modelling, where it is emphasized to fit a wide variety of parametric models (either proportional hazards or accelerated failure time), to obtain the `best fitting' model \citep{Latimer2013a}. Often, extrapolation is required to calculate survival across a lifetime horizon, and hence parametric and flexible approaches are needed. Of course, extrapolation is fraught with dangers, and arguably should only be attempted in the presence of appropriate external data \citep{Andersson2013}.

To our knowledge, the most flexible fully parametric AFT model is the generalized F distribution, a four parameter distribution described by \cite{Cox2008}, which often suffers from convergence problems. This contains the more widely used (due to availability of software) generalized gamma as a special case \citep{Cox2007}. Many authors have compared and contrasted accelerated failure time models with the more commonly used proportional hazards metric \citep{Kay2002,Orbe2002}. \cite{Lambert2004} developed a mixture AFT model with frailties, where a short term hazard component was modelled with a Gompertz distribution, and the long term hazard component could be any of the standard parametric AFT models. There have been several efforts to develop smooth accelerated failure time models, including mixtures of normal densities \citep{Komarek_Lesaffre_Hilton_2005}, kernel smoothed densities \citep{Zeng_Lin_2007} and seminonparametric densities \citep{Zhang_Davidian_2008}. A software implementation of the mixture of normal densities has received modest attention. \cite{Rubio2019} recently proposed a general hazards-based model, utilising the exponentiated-Weibull model to model the baseline function.

Within a proportional hazards metric, the Royston-Parmar flexible parametric model has grown in popularity in recent years, with a number of extensions and developments being proposed \citep{FPMbook,Liu2016}. The fundamental strength of the model is to use restricted cubic splines to model the underlying baseline function (regardless of scale), and any time-dependent effects. However, there are known limitations with models based on hazard ratios, where the hazard ratios are not collapsible across covariates not associated the exposure of interest \citep{Martinussen_Vansteelandt_2013}, while AFTs are known to be robust to omitted covariates \citep{Hougaard1999}. Together, this motivates the incorporation of a flexible framework into an accelerated failure time paradigm, which we consider in this article. 

AFT models make the assumption of a constant acceleration factor, i.e. the effect of a covariate remains the same across follow-up time, similar to the proportional hazards assumption. Clearly this assumption is open to violation. This motivates the relaxation of the constant acceleration factor to allow time-dependency, similarly to modelling of non-proportional hazards. This has been described within a generalised gamma AFT model by \cite{Cox2007}. In this article, we further relax the constant acceleration factor assumption, within the flexible parametric AFT model, by using restricted cubic splines.

The paper is organised as follows. In Section~\ref{sec:theory}, we first show that an accelerated failure time model has some desirable properties, including collapsibility, that are not exhibited by a proportional hazards model. In Section \ref{sec:methods}, we derive the proposed model framework and describe the estimation process within a likelihood framework. In Section \ref{sec:sim} we conduct a simulation study to evaluate the finite sample performance of the proposed model under complex scenarios, comparing to standard parametric AFT models. In Section \ref{sec:data} we illustrate the model using data from the England and Wales breast cancer registry. Finally, in Section \ref{sec:disc} we conclude the paper with a discussion.

\section{Causal interpretation of the accelerated failure time model} \label{sec:theory}

\begin{figure}[!ht]
  \centering
    \begin{tikzpicture}[->,bend angle=20,semithick,>=stealth']
      \matrix [matrix of nodes,row sep=7mm, column sep=10mm]
      {
        |(X)| $X$ & |(C)| $C$ \\
        & |(T)| $T$ & |(Y)| $(Y,\Delta)$ \\
        |(Z)| $Z$ \\
      };
      \begin{scope}[every node/.style={midway,auto}]
        \draw (X) to node[anchor=south] {} (T);
        \draw (Z) to node[anchor=south] {} (T);
        \draw (T) to node[anchor=north] {} (Y);
        \draw[<->] (X) to [bend right=60] node {} (Z);
        \draw (C) to node[anchor=north] {} (Y);
      \end{scope}
    \end{tikzpicture}  
  \caption{Causal diagram for the motivating example, with censoring variable $C$ being independent from $X$, $Z$ and $T$.}
  \label{fig:causal}
\end{figure}
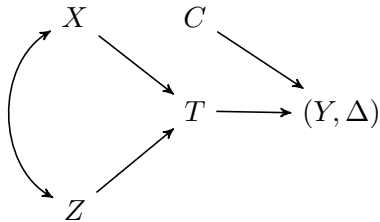

\cite{Hougaard1999} provided an informal description of how AFT models are robust to omitted covariates. We now provide a more formal development for the collapsibility of the acceleration factor for AFT models.
Consider a model with two covariates $X$ and $Z$, with an event time $T$, with regression parameters $\beta_X$ and $\beta_Z$ and linear predictor $\beta_X x + \beta_Z z$. Assume that the censoring variable $C$ is independent of $X$, $Z$ and $T$, and that the time process is observed by the tuple $(Y=\min(T,C),\Delta=I(T\leq C))$; the associated causal diagram is given in Figure~\ref{fig:causal}. 

Following \cite{Martinussen_Vansteelandt_2013}, we define the marginal unadjusted effect for binary $X=1$ compared with $X=0$ at time $t$ for a contrast function $g$ and a prediction function $\psi$ as $\mathcal{T}_m(t)=g(\psi(t|x=1),\psi(t|x=0))$. The marginal exposure effect (causal effect) is defined as $\mathcal{T}(t)=g(\psi(t|\hat{x}=1),\psi(t|\hat{x}=0)$, where $\hat{x}=x$ is the \emph{do} operator, which can be conceptualised as the population value that would be realised if $X$ were uniformly set to $x$. 

\subsection{Proportional hazard model}
\cite{Martinussen_Vansteelandt_2013} considered a proportional hazards model $h(t|x,z)=h_0(t)\exp(\beta_X x + \beta_Z z)$. The marginal hazard conditional on survival has previously been shown to be
\begin{align*}
  E_Z(h(t|x,Z)|T>t) &= \frac{E_Z(h(t|x,Z) S(t|x,Z))}{E_Z(S(t|x,Z))} \\
                    &= \frac{h_0(t)\exp(\beta_X x) E_Z(\exp(\beta_Z Z) S(t|x,Z))}{E_Z(S(t|x,Z))}
\end{align*}
For the marginal (causal) effect for the log-hazard ratio, we define $\psi(t|\hat{x}=x)=E_Z(h(t|\hat{x}=x,Z)|T>t)$ and $g(a,b)=\log(a/b)$, then the marginal (causal) effect is
\begin{align}
  \mathcal{T}(t) &= \log(E_Z(h(t|\hat{x}=1,Z)|T>t))-\log(E_Z(h(t|\hat{x}=0,Z)|T>t)) \nonumber \\
                 &= \beta_X +
                   \log\left(\frac{E_Z(\exp(\beta_Z Z) S(t|\hat{x}=1,Z))}{E_Z(S(t|\hat{x}=1,Z))}\right) - 
                   \log\left(\frac{E_Z(\exp(\beta_Z Z) S(t|\hat{x}=0,Z))}{E_Z(S(t|\hat{x}=0,Z))}\right)  \label{eq:causal}
\end{align}
For a marginal unadjusted effect, we define $\psi(t|x)=E_Z(h(t|x,Z)|T>t,X=x)$. Then 
\begin{align}
  \mathcal{T}_m(t) &= \log(E_Z(h(t|x=1,Z)|T>t,X=1))-\log(E_Z(h(t|x=0,Z)|T>t,X=0)) \nonumber \\
                   &= \beta_X +
                     \log\left(\frac{E_Z(\exp(\beta_Z Z) S(t|x=1,Z) | X=1)}{E_Z(S(t|x=1,Z)|X=1)}\right) - 
                     \log\left(\frac{E_Z(\exp(\beta_Z Z) S(t|x=0,Z) | X=0)}{E_Z(S(t|x=0,Z) | X=0)}\right) \label{eq:unadjusted}
\end{align}
These expressions show that $\beta_X$ is a biased estimator of both the marginal causal effect and the marginal unadjusted effect. We can use Equation~(\ref{eq:causal}) to estimate the causal effect $\hat{\mathcal{T}}(t)$ from a model fit incorporating both $X$ and $Z$, and use Equations~(\ref{eq:causal}) and (\ref{eq:unadjusted}) to calculate the confounding bias $\mathcal{T}(t)-\mathcal{T}_m(t)$.
We can also use Equation~(\ref{eq:unadjusted}) to calculate the bias in the unadjusted estimate for $\beta_X$ when $Z$ is not modelled compared with modelling both $X$ and $Z$ (that is, $\mathcal{T}_m(t)-\beta_X$) when $\beta_Z \neq 0$ and $X$ and $Z$ are associated.

\subsection{Accelerated failure time model}

Now define an accelerated failure time model $\log(T) = \alpha -(\beta_X x + \beta_Z z) + \epsilon$, where $\alpha=-\beta_0$ and $E(\epsilon)=0$. For the marginal (causal) effect for the mean time to event comparing $X=1$ with $X=0$, define $\psi(t|x)=E_Z(\log(T)|\hat{x}=x,Z)$ and $g(a,b)=a-b$, such that
\begin{align*}
  \mathcal{T} &= E_Z(\log(T)|\hat{x}=1,Z)-E_Z(\log(T)|\hat{x}=0,Z) = -\beta_X
\end{align*}
which is unbiased and indicates \emph{collapsibility} of the acceleration factor. For the marginal unadjusted effect, let $\psi(t|x)=E_Z(\log(T)|X=x,Z)$, so that
\begin{align*}
  \mathcal{T}_m = -\beta_X - \beta_Z(E_Z(Z|X=1)-E_Z(Z|X=0))
\end{align*}
which will be biased if, again, $\beta_Z\neq 0$ and $X$ and $Z$ are associated. We can extend this finding to a time-dependent accelerated failure time model $S(t|x,z)=S_0\left(\int_0^t \exp(\eta(u|x,z)) du\right)$ for a time-varying acceleration factor $\eta(u|x,z)=-\beta_X(u) x - \beta_Z(u) z$ and for a baseline survival function $S_0(t)$.
The marginal value for $\eta(t|x,Z)$ conditional on survival is
\begin{align*}
  E_Z(\eta(t|x,Z)| T>t) &= -\beta_X(t) x - \beta_Z(t) E_Z(Z|x)
\end{align*}
then the marginal (causal) effect comparing $X=1$ with $X=0$ at time $t$ is
\begin{align*}
  \mathcal{T}(t) &= E_Z(\eta(t|\hat{x}=1,Z)) - E_Z(\eta(t|\hat{x}=0,Z)) = -\beta_X(t)
\end{align*}
For the marginal unadjusted effect, let $\psi(t|x)=E_Z(\eta(t|X=x,Z))$, and then
\begin{align*}
  \mathcal{T}_m(t) &= -\beta_X(t) - \beta_Z(t) (E_Z(Z|X=1) - E_Z(Z|X=0))
\end{align*}

Pleasantly, if $E_Z(Z|X=1) - E_Z(Z|X=0)$ is small, then $\mathcal{T}_m(t) \approx -\beta_X(t)$ and the AFT can be shown to be robust to omitted covariates.

\section{A general parametric accelerated failure time model}
\label{sec:methods}
Continuing with the notation defined in the previous section, an accelerated failure time model, conditional on a set of explanatory variables, $\boldsymbol{X}$, can be written in the form of the survival function,
\begin{equation}
	S(t | \boldsymbol{X}) = S_{0} (t\, \phi(\boldsymbol{X};\,\boldsymbol\beta)) \nonumber
	\label{eqn:surv}
\end{equation}
where often
\begin{equation}
	\phi(\boldsymbol{X};\,\boldsymbol\beta) = \exp(-\boldsymbol{X} \boldsymbol{\beta})
	\label{eqn:covs1}
\end{equation}
We can also specify an AFT model in terms of the cumulative hazard function
\begin{equation}
	H(t | \boldsymbol{X}) = H_{0} (t\, \phi(\boldsymbol{X};\,\boldsymbol\beta))
	\label{eqn:ch}
\end{equation}
In essence, we can specify any parametric function for Equation (\ref{eqn:ch}), subject to the appropriate constraints that the function remains positive for all $t>0$ and is monotonically increasing as $t \rightarrow \infty$. In this article, we concentrate on a highly flexible way of specifying a parametric AFT, using restricted cubic splines as our basis functions \cite{Durrleman1989}.

Similarly to \cite{Royston2002}, we begin with the log cumulative hazard function of the Weibull distribution,
\begin{equation}
	\log H(t|\lambda,\gamma) = \log (\lambda) + \gamma \log (t)  \nonumber
\end{equation}
Instead of incorporating covariates into the linear predictor of the $\log(\lambda)$ component, as in \cite{Royston2002}, here we incorporate them as a multiplicative effect on $t$,
\begin{equation}
	\log H(t | \boldsymbol{X},\lambda,\gamma,\boldsymbol\beta) = \log (\lambda) + \gamma \log (t\, \phi(\boldsymbol{X};\,\boldsymbol\beta)) \nonumber
\end{equation}
where $\phi(\boldsymbol{X};\,\boldsymbol\beta)$ is defined in Equation (\ref{eqn:covs1}). Now we can incorporate the desired flexibility, expanding $\log (t\, \phi(\boldsymbol{X};\,\boldsymbol\beta))$ into restricted cubic spline basis. For simplicity, letting $u = \log (t\, \phi(\boldsymbol{X};\,\boldsymbol\beta))$, our spline function is defined as
\begin{equation}
	s(u|\boldsymbol{\gamma},\boldsymbol{k_{0}}) = \gamma_{0} + \gamma_{1}v_{1}(u,\boldsymbol{k_{0}}) + \gamma_{2}v_{2}(u,\boldsymbol{k_{0}}) + \dots + \gamma_{m+1}v_{m+1}(u,\boldsymbol{k_{0}}) \nonumber
\end{equation}
where $\boldsymbol{k}_{0}$ is a vector of knot locations with parameter vector $\boldsymbol{\gamma}$, and derived variables $v_{j}$ (known as the basis functions). For a truncated power basis, the $v_j$ are defined as
\begin{align}
	v_{1}(u,\boldsymbol{k_{0}}) &= u  \nonumber\\
	v_{j}(u,\boldsymbol{k_{0}}) &= (u-k_{j})^{3}_{+} - \lambda_{j}(u-k_{\text{min}})^{3}_{+} - (1-\lambda_{j})(u-k_{\text{max}})^{3}_{+} \nonumber
\end{align}
where for $j=2,\dots,m+1$, $(u-k_{j})^{3}_{+}$ is equal to $(u-k_{j})^{3}$ if the value is positive and 0 otherwise, and
\begin{equation}
	\lambda_{j} = \frac{k_{\text{max}}-k_{j}}{k_{\text{max}}-k_{\text{min}}} \nonumber
\end{equation}
Alternatively, the $v_j(u,\boldsymbol{k_{0}})$ can be calculated using a B-spline basis with a matrix projection at the boundary knots, as per the \texttt{ns} function in R. Given one of these bases, our flexible parametric AFT model can be defined as
\[
	\log H(t | \boldsymbol{X}) = s(\log(t\, \phi(\boldsymbol{X};\,\boldsymbol\beta)) | \boldsymbol{\gamma},\boldsymbol{k}_{0})
\]
Usually, knot locations are calculated based on quantiles of the distribution of the variable being transformed into splines, in this case $\log(t\, \phi(\boldsymbol{X};\,\boldsymbol\beta))$, also restricted to those observations which are uncensored. 

\subsection{Likelihood and estimation}

We define the likelihood in terms of the hazard and survival functions. The hazard function can be written as follows
\begin{align}
	h(t|\boldsymbol X) &= \frac{d}{dt} H(t|\boldsymbol X)  \nonumber \\
			&= \exp[s(\log(t\,\phi(\boldsymbol{X};\,\boldsymbol\beta))|\boldsymbol{\gamma},\boldsymbol{k}_{0})] \times \frac{d}{d t} [ s(\log(t\,\phi(\boldsymbol{X};\,\boldsymbol\beta))|\boldsymbol{\gamma},\boldsymbol{k}_{0}) ]  \nonumber \\
			&= \exp[s(\log(t\,\phi(\boldsymbol{X};\,\boldsymbol\beta))|\boldsymbol{\gamma},\boldsymbol{k}_{0})] \times \frac{1}{t} \times s^{\prime}(\log(t\,\phi(\boldsymbol{X};\,\boldsymbol\beta))|\boldsymbol{\gamma},\boldsymbol{k}_{0}) \nonumber
\end{align}
where $s^{\prime}(x) = \frac{d}{dx}s(x)$. The survival function is defined as
\begin{equation}
	S(t|\boldsymbol X) = \exp[-\exp\{s(\log(t\,\phi(\boldsymbol{X};\,\boldsymbol\beta))|\boldsymbol{\gamma},\boldsymbol{k}_{0})\}] \nonumber
\end{equation}
We can therefore define our log likelihood for the $i^{th}$ patient, allowing for delayed entry, as
\begin{align}
	l_{i} &= d_{i} \log h(y_{i}) + \log S(y_{i}) - \log S(t_{0i})  \nonumber \\
	&= d_{i} \times \left[ s(\log(y_{i}\,\phi(\boldsymbol{X};\,\boldsymbol\beta))|\boldsymbol{\gamma},\boldsymbol{k}_{0}) -\log(y_{i}) + \log(s^{\prime}(\log(y_{i}\,\phi(\boldsymbol{X};\,\boldsymbol\beta))|\boldsymbol{\gamma},\boldsymbol{k}_{0}))  \right] \nonumber \\
&\qquad \quad	-\exp \{s(\log(y_{i}\,\phi(\boldsymbol{X};\,\boldsymbol\beta))|\boldsymbol{\gamma},\boldsymbol{k}_{0}) \} + \exp \{s(\log(t_{0i}\,\phi(\boldsymbol{X};\,\boldsymbol\beta))|\boldsymbol{\gamma},\boldsymbol{k}_{0})\}
\label{eqn:ll}
\end{align}
We maximise Equation (\ref{eqn:ll}) using Newton-Raphson based optimisation \citep{Gouldml}, with analytic score and Hessian.

\subsection{Time-dependent acceleration factors}

Following \cite{Cox1984} and \cite{Hougaard1999}, we have the survival function of a time-dependent AFT model, such that
\[
  S(t) = S_{0} \left( \int_{0}^{t} \eta(\boldsymbol X, u;\,\boldsymbol\beta) \text{ d} u \right)
\]
where $\eta(\boldsymbol X, u;\,\boldsymbol\beta)$ is the time-varying acceleration factor at time $u$ and the baseline survival is $S_0(t)=\exp(-\exp(s(\log(t)|\boldsymbol{\gamma},\boldsymbol{k}_{0})))$. Within our flexible parametric framework, we can avoid the integration by directly modelling on the cumulative scale, such that
\[
  S_{0} \left( \int_{0}^{t} \eta(\boldsymbol X, u;\,\boldsymbol\beta) \text{ d} u \right) = S_{0}(t \times \phi(\boldsymbol X, t;\,\boldsymbol\beta))
\]
Since we are on a cumulative scale, to recover the directly interpretable time-dependent acceleration factor, $\eta(\boldsymbol X, t;\,\boldsymbol\beta)$, we derive the following relationship,
\begin{align}
  \int_{0}^{t} \eta(\boldsymbol X, u;\,\boldsymbol\beta) \text{ d} u &= t \times \phi(\boldsymbol{X}, t;\,\boldsymbol\beta) \nonumber \\
  \implies \eta(\boldsymbol X, t;\,\boldsymbol\beta)  &= \frac{d}{d t} \left[ t \phi(\boldsymbol{X},t;\,\boldsymbol\beta) \right] \nonumber  \\
   &= \phi(\boldsymbol{X},t;\,\boldsymbol\beta) + t\,\frac{d}{d t} \phi(\boldsymbol{X}, t;\,\boldsymbol\beta)
\end{align}
which gives a rather convenient formula for the time-dependent acceleration factor in terms of its cumulative. We can arguably use any continuous function to capture simple and complex time-dependent acceleration factors. The hazard function is defined as
\begin{align}
	h(t| \boldsymbol X) 	&= \exp[s(\log(t \phi(\boldsymbol{X},t))|\boldsymbol{\gamma},\boldsymbol{k}_{0})] \times \frac{d }{d t} [ s(\log(t \phi(\boldsymbol{X},t))|\boldsymbol{\gamma},\boldsymbol{k}_{0}) ] \nonumber  \\
				&= \exp[s(\log(t \phi(\boldsymbol{X},t))|\boldsymbol{\gamma},\boldsymbol{k}_{0})] \times \frac{t \frac{d  \phi(\boldsymbol{X},t)}{d t} + \phi(\boldsymbol{X},t) }{t \phi(\boldsymbol{X},t)} \times s^{\prime}(\log(t \phi(\boldsymbol{X},t))|\boldsymbol{\gamma},\boldsymbol{k}_{0})
				\label{eqn:tdehaz1}
\end{align}

Our form of choice continues the use of restricted cubic splines. A common case is to use a minus log link for the linear predictor, such that
\begin{equation}
	  \phi(\boldsymbol{X},t;\,\boldsymbol\beta) = \exp\left( -\boldsymbol{X} \boldsymbol{\beta} - \sum_{p=1}^{P} x_{p} s(\log(t)| \boldsymbol{\gamma}_{p}, \boldsymbol{k}_{p})\right)
	\label{eqn:tde}
\end{equation}
where for the $p^{th}$ time-dependent effect, with $p = \{1,\dots,P\}$, we have $x_{p}$, the $p^{th}$ covariate, multiplied by some spline function of log time, $s(\log(t)| \boldsymbol{\gamma}_{p},\boldsymbol{k}_{p})$, with knot location vector, $\boldsymbol{k}_{p}$, and coefficient vector, $\boldsymbol{\gamma}_{p}$.

Now
\begin{align}
	 \frac{d \phi(\boldsymbol{X},t;\,\boldsymbol\beta)}{d t} &= \frac{d}{d t} \exp\left( -\boldsymbol{X} \boldsymbol{\beta} - \sum_{p=1}^{P} x_{ip} s(\log(t)| \boldsymbol{\gamma}_{p}, \boldsymbol{k}_{p})\right) \nonumber  \\
	 &= \phi(\boldsymbol{X},t;\,\boldsymbol\beta) \times \frac{d}{d t} \left[ -\boldsymbol{X} \boldsymbol{\beta} - \sum_{p=1}^{P} x_{p} s(\log(t)| \boldsymbol{\gamma}_{p}, \boldsymbol{k}_{p}) \right] \nonumber  \\
	 &= \phi(\boldsymbol{X},t;\,\boldsymbol\beta) \times \left[ -\sum_{p=1}^{P} x_{ip} \frac{d \log(t)}{d t} \frac{d}{d \log(t)} s(\log(t)| \boldsymbol{\gamma}_{p}, \boldsymbol{k}_{p}) \right] \nonumber \\
	 &= \frac{\phi(\boldsymbol{X},t;\,\boldsymbol\beta)}{t} \times \left[ -\sum_{p=1}^{P} x_{p} s^{\prime}(\log(t)| \boldsymbol{\gamma}_{p}, \boldsymbol{k}_{p}) \right]
	 \label{eqn:tdehaz2}
\end{align}
Substituting Equation (\ref{eqn:tdehaz2}) into (\ref{eqn:tdehaz1})
\begin{align}
	h(t|\boldsymbol X) 	&= \exp[s(\log(t \phi(\boldsymbol{X},t;\,\boldsymbol\beta))|\boldsymbol{\gamma},\boldsymbol{k}_{0})] \times s^{\prime}(\log(\phi(\boldsymbol{X},t;\,\boldsymbol\beta))|\boldsymbol{\gamma},\boldsymbol{k}_{0}) \times \frac{1}{t} \times \left[1 - \sum_{p=1}^{P} x_{p} s^{\prime}(\log(t)| \boldsymbol{\gamma}_{p}, \boldsymbol{k}_{p})\right]
				\label{eqn:tdehaz3}
\end{align}
with survival function
\begin{equation}
	S(t|\boldsymbol X) = \exp[-\exp\{s(\log(t\phi(\boldsymbol{X},t;\,\boldsymbol\beta))|\boldsymbol{\gamma},\boldsymbol{k}_{0})\}]
	\label{eqn:tdesurv1}
\end{equation}
Equations (\ref{eqn:tdehaz3}) and (\ref{eqn:tdesurv1}) can then be substituted into Equation (\ref{eqn:ll}) to maximise the log likelihood. Analytic scores and Hessian elements can also be derived.

\section{Simulations}
\label{sec:sim}

\subsection{Causal inference}

We first simulate under Figure~\ref{fig:causal}. Assume that $X$ is Bernoulli, $Z$ is normal, $T$ is exponential and $C$ is uniform. Specifically, let $X\sim\text{Bernoulli}(0.5), Z\sim\text{Normal}(0,2^2), T\sim\text{Exponential}(\exp(\beta_0+\beta_XX+\beta_ZZ)), \beta_0=-5, \beta_X=\beta_Z=1, C\sim\text{Uniform}(0,10)$ and $(Y,\Delta)=(\min(T,C),T<C)$. Let $(x,z,y,\delta)$ be realisations for $(X,Z,Y,\Delta)$. These data can be modelled using both proportional hazards and accelerated failure time models. We fit models for $(y,\delta)$ with both $x$ and $z$ as linear and additive covariates and with only $x$ as a covariate. We model using Poisson regression, Cox regression and our smooth AFT with 3 degrees of freedom (see Table~\ref{tab:simcausal}). Note that the estimated $\beta_X$'s have opposite signs for the AFT models compared with the proportional hazards models (Poisson and Cox regression). As a reminder, an exponential AFT would estimate $-\beta_X$ as per the Poisson regression. We find that all of the models are unbiased when both covariates are included ($|E(\hat\beta_X)|=1$). 

When $X$ and $Z$ are independent, then the effect of $X$ is not confounded by $Z$ and $\mathcal{T}_m(t)=\mathcal{T}(t)$. For the AFT models, assuming that we have captured the baseline distribution, then $E_Z(Z|X=1)=E_Z(Z|X=0)$ and $\mathcal{T}_m(t)=-\beta_X(t)$. However, for the proportional hazards models with both covariates, the marginal (causal) effect is attenuated for increasing time. This pattern of attenuation is well recognised from the context of frailty models (e.g. \cite{Aalen_Borgan_Gjessing_2008}). Moreover, when the covariate $z$ is not included and $X$ and $Z$ are correlated, then the first three models are more biased, while the smooth AFT is less biased.

\begin{table}[!ht]
\caption{Simulation results for exponentially distributed data with $\beta_X=1$, with $n=10^4$ observations per simulation and 300 simulation sets. The regression models assume that both covariates are modelled ($(y,\delta) \sim x+z$) or that only the $x$ covariate is modelled ($(y,\delta)\sim x$). The expectations for the estimated $\beta_X$ and their standard errors are over the simulation sets.}
\label{tab:simcausal}
\begin{center}
  \begin{tabular}{clrrrrr}
    && \multicolumn{2}{c}{$(y,\delta) \sim x+z$} && \multicolumn{2}{c}{$(y,\delta)\sim x$} \\
    \cline{3-4}\cline{6-7}
$\text{Corr}(X,Z)$ & Model & $E(\hat\beta_X)$ & $E(se(\hat\beta_X))$ && $E(\hat\beta_X)$ & $E(se(\hat\beta_X))$\\
                              \hline
0 & Poisson regression  & 0.998 & 0.054 && 0.679 & 0.053\\
& Cox regression  & 0.999 & 0.054 && 0.663 & 0.053\\
& Smooth AFT & -0.998 & 0.055 && -0.962 & 0.079 \\
                              \hline
0.1 & Poisson regression  & 0.998 & 0.054 && 0.901 & 0.054\\
& Cox regression  & 0.999 & 0.055 && 0.880 & 0.054\\
& Smooth AFT & -0.998 & 0.056 && -1.278 & 0.083\\
                              \hline
-0.1 & Poisson regression  & 0.998 & 0.053 && 0.464 & 0.052\\
& Cox regression  & 0.999 & 0.054 && 0.454 & 0.052\\
& Smooth AFT & -0.998 & 0.055 && -0.657 & 0.077 
\end{tabular}
\end{center}
\end{table}

\subsection{Other simulations}
In this section we conduct a simulation study to assess the ability of the flexible parametric AFT model to capture complex, biologically plausible, baseline functions, and subsequently the impact on estimates of acceleration factors and survival probabilities, when misspecifying the baseline. We also compare the newly proposed flexible AFT to existing parametric models, including the Weibull, generalized gamma, and generalized F. The Weibull and generalized gamma models are available in the {\tt streg} command in Stata, and we implement the generalized F in Stata, following \cite{Cox2008}.

\begin{figure}[!ht]
	\centering
	\includegraphics{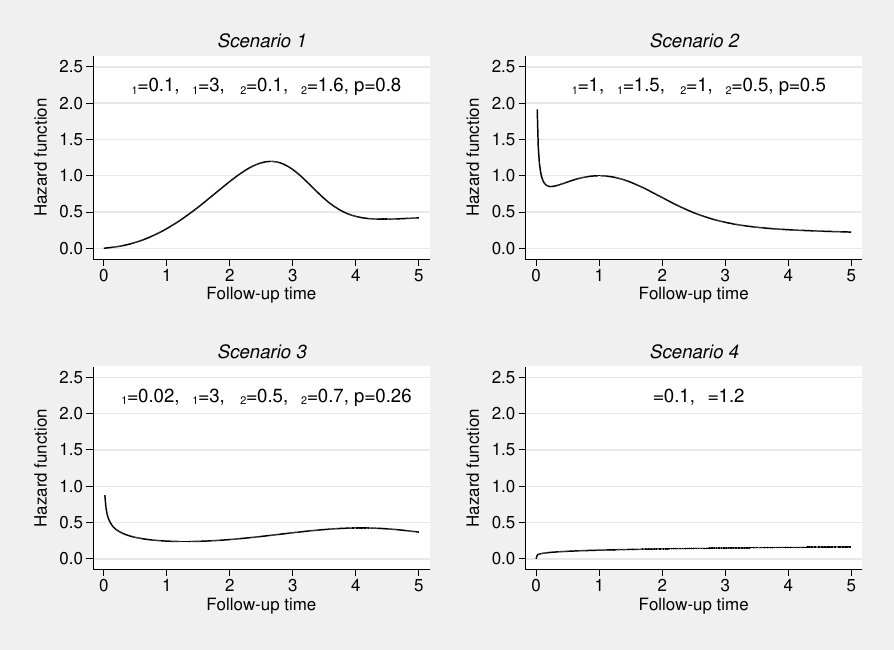}
	\caption{Baseline hazard functions for the simulation scenarios.}
	\label{fig:bhaz}
\end{figure}

In all simulations, we use a range of two-component mixture Weibull baseline hazard functions, and also a standard Weibull, to generate complex, realistic scenarios \citep{CrowtherSurvsim}. When fitting the flexible parametric models, we are therefore not fitting the `true' model, but investigating how well the spline approximations can do \citep{Rutherford2013}. The baseline survival function for a two-component mixture Weibull is defined as follows:
\begin{equation}
	S_{0}(t) = p \exp (-\lambda_{1}t^{\gamma_{1}}) + (1-p) \exp(-\lambda_{2}t^{\gamma_{2}})
	\label{eqn:mixweib}
\end{equation}
We choose four different baseline hazard functions, representing clinically plausible functions \citep{FPMbook,Murtaugh1994}. The four assumed baseline hazard functions are shown in Figure \ref{fig:bhaz}. Scenarios 1 to 3 come from mixture Weibull functions defined in Equation (\ref{eqn:mixweib}), with Scenario 4 a standard Weibull function.
With our baseline functions defined, we can choose to simulate under an accelerated failure time framework, or under proportional hazards, using the general survival simulation framework developed by \cite{CrowtherSurvsim}.

Consider a binary treatment group variable, $X$, with a log acceleration factor, $\beta$. We can simulate accelerated failure time data from the following,
\begin{align}
	S(t| X) &= p \exp (-\lambda_{1}(t e^{-X\beta})^{\gamma_{1}}) + (1-p) \exp(-\lambda_{2}(t e^{-X\beta})^{\gamma_{2}}) \nonumber \\
	&= p \exp (-\lambda_{1}t^{\gamma_{1}} e^{-X\beta\gamma_{1}}) + (1-p) \exp(-\lambda_{2}t^{\gamma_{2}} e^{-X\beta\gamma_{2}})
	\label{eqn:mwaf}
\end{align}
For each scenario, we assume a log AF of $\beta=-0.5$, or $\beta=0.5$. This results in 8 scenarios in total.

To each simulated dataset, we apply a Weibull AFT model, a generalised gamma AFT model, a generalized F AFT model and the proposed flexible parametric AFT model with 2 to 9 degrees of freedom. We do not fit a flexible parametric acceleration failure time model with 1 degree of freedom, as this is equivalent to a Weibull AFT model. Each simulation scenario is repeated 250 times with 1000 observations in each dataset. We set a maximum follow-up time of 5 years. The following average survival probability at 5 years was observed in each scenario; $\bar{S}(5)=0.03$ in scenario 1 and $\beta=-0.5$, $\bar{S}(5)=0.106$ in scenario 1 and $\beta=0.5$, $\bar{S}(5)=0.040$ in scenario 2 and $\beta=-0.5$, $\bar{S}(5)=0.071$ in scenario 1 and $\beta=0.5$, $\bar{S}(5)=0.131$ in scenario 3 and $\beta=-0.5$, $\bar{S}(5)=0.289$ in scenario 3 and $\beta=0.5$, $\bar{S}(5)=0.393$ in scenario 4 and $\beta=-0.5$, $\bar{S}(5)=0.592$ in scenario 4 and $\beta=0.5$.

We monitor estimates of $\beta$ from all models, and estimates of the survival probability at 1, 2, 3, 4, and 5 years, in both treatment groups. Survival was monitored on the $\log [-\log()]$ scale, with standard errors calculated using the delta method. We also monitor values of the AIC and BIC.

\subsection{Simulation results}

Results are presented in Table \ref{tab:sim} for all 8 scenarios. We present bias, percentage bias, and coverage for the estimates of the log acceleration factor from all AFT models. We further present the median rank in terms of best fitting model based on either the AIC or BIC, for all models fitted. Finally, in Tables \ref{tab:simpos0} to \ref{tab:simneg1} we present bias, percentage bias, and coverage for estimates of the survival probability at 1, 2, 3, 4, and 5 years, for the four scenarios, when $X=0$ and $\beta=0.5$, $X=1$ and $\beta=0.5$, $X=0$ and $\beta=-0.5$, $X=1$ and $\beta=-0.5$, respectively, with all estimates are on the $\log \left\{-\log[S(t)]\right\}$ scale.

\begin{sidewaystable}[!ht]
\caption{Simulation results}
\label{tab:sim}
\begin{center}
\scalebox{0.6}{
\begin{tabular}{c c r r r r r r r r r r r r r r r r r r r r r r r r} \hline
\multirow{2}{*}{True log(AF)} & \multirow{2}{*}{Model} & \multicolumn{6}{c}{Scenario 1} & \multicolumn{6}{c}{Scenario 2} & \multicolumn{6}{c}{Scenario 3} & \multicolumn{6}{c}{Scenario 4}  \\
 & & Bias & \% Bias & Cov. & AIC & BIC & \# Conv. & Bias & \% Bias & Cov. & AIC & BIC & \# Conv.  & Bias & \% Bias & Cov. & AIC & BIC & \# Conv. & Bias & \% Bias & Cov. & AIC & BIC & \# Conv.  \\
\hline 
\multirow{11}{*}{0.5} & Weibull & -0.082 & -16.4 & 20.8 & 11 & 11 & 250 & -0.049 & -9.8 & 92.0 & 11 & 10 & 250 & 0.138 & 27.6 & 70.8 & 11 & 9 & 250 & 0.003 &  0.6 & 96.4 & 1 & 1 & 250 \\
& Gamma & -0.033 & -6.6 & 84.0 & 9 & 9 & 250 & -0.074 & -14.8 & 84.0 & 10 & 10 & 250 & 0.032 &  6.4 & 70.4 & 5 & 2 & 181 & 0.005 &  1.0 & 89.6 & 3 & 3 & 235 \\
& GenF & 0.001 &  0.2 & 92.8 & 7 & 1 & 247 & 0.000 &  0.0 & 73.6 & 6 & 1 & 191 & 0.032 &  6.4 & 41.2 & 5 & 3 & 109 & 0.021 &  4.2 & 51.6 & 5 & 5 & 140 \\
& FPAFT-df=2 & -0.011 & -2.2 & 94.0 & 10 & 10 & 250 & -0.076 & -15.2 & 82.0 & 9 & 8 & 250 & 0.107 & 21.4 & 78.0 & 10 & 7 & 250 & 0.005 &  1.0 & 95.6 & 3 & 3 & 250 \\
& FPAFT-df=3 & 0.021 &  4.2 & 88.4 & 8 & 3 & 250 & 0.019 &  3.8 & 96.0 & 8 & 7 & 250 & 0.024 &  4.8 & 95.6 & 2 & 2 & 250 & 0.006 &  1.2 & 94.0 & 5 & 4 & 250 \\
& FPAFT-df=4 & 0.011 &  2.2 & 92.4 & 3.5 & 2 & 250 & 0.023 &  4.6 & 91.6 & 6 & 2 & 250 & 0.003 &  0.6 & 96.0 & 3 & 4 & 250 & 0.005 &  1.0 & 92.4 & 6 & 6 & 249 \\
& FPAFT-df=5 & 0.008 &  1.6 & 94.0 & 2 & 4 & 250 & 0.019 &  3.8 & 92.0 & 4 & 3 & 250 & 0.006 &  1.2 & 95.6 & 4 & 5 & 250 & 0.005 &  1.0 & 88.8 & 7 & 7 & 245 \\
& FPAFT-df=6 & 0.005 &  1.0 & 92.4 & 3 & 5 & 250 & 0.014 &  2.8 & 92.4 & 3 & 4 & 250 & 0.008 &  1.6 & 95.2 & 6 & 6 & 248 & 0.008 &  1.6 & 85.2 & 8 & 8 & 238 \\
& FPAFT-df=7 & 0.004 &  0.8 & 91.6 & 4 & 6 & 246 & 0.011 &  2.2 & 92.8 & 3 & 5 & 248 & 0.007 &  1.4 & 94.8 & 7 & 8 & 250 & 0.013 &  2.6 & 73.6 & 9 & 9 & 226 \\
& FPAFT-df=8 & 0.004 &  0.8 & 87.2 & 5 & 7 & 239 & 0.009 &  1.8 & 90.4 & 4 & 6 & 245 & 0.007 &  1.4 & 90.8 & 7 & 9 & 244 & 0.009 &  1.8 & 68.8 & 10 & 10 & 225 \\
& FPAFT-df=9 & 0.003 &  0.6 & 80.4 & 6 & 8 & 227 & 0.009 &  1.8 & 86.8 & 5 & 7 & 243 & 0.008 &  1.6 & 85.2 & 9 & 11 & 236 & 0.012 &  2.4 & 60.0 & 11 & 11 & 217 \\
\hline 
\multirow{11}{*}{-0.5} & Weibull & 0.057 & -11.4 & 59.6 & 11 & 11 & 250 & 0.041 & -8.2 & 93.2 & 10 & 9 & 250 & 0.008 & -1.6 & 96.0 & 10 & 10 & 250 & 0.003 & -0.6 & 96.4 & 1 & 1 & 250 \\
& Gamma & 0.011 & -2.2 & 93.2 & 10 & 10 & 248 & 0.047 & -9.4 & 92.0 & 10 & 11 & 250 & 0.077 & -15.4 & 80.0 & 8 & 5 & 250 & 0.001 & -0.2 & 95.6 & 3 & 3 & 250 \\
& GenF & 0.001 & -0.2 & 92.8 & 3 & 2 & 248 & 0.001 & -0.2 & 70.0 & 7 & 1 & 183 & -0.002 &  0.4 & 41.6 & 2 & 1 & 106 & -0.011 &  2.2 & 72.8 & 5 & 5 & 188 \\
& FPAFT-df=2 & 0.026 & -5.2 & 88.0 & 9 & 9 & 250 & 0.040 & -8.0 & 92.0 & 9 & 9 & 250 & 0.038 & -7.6 & 94.8 & 9 & 7 & 250 & 0.001 & -0.2 & 96.0 & 3 & 3 & 250 \\
& FPAFT-df=3 & -0.001 &  0.2 & 94.0 & 2 & 1 & 250 & 0.107 & -21.4 & 79.6 & 8 & 7 & 250 & 0.042 & -8.4 & 91.6 & 7 & 5 & 250 & 0.000 &  0.0 & 96.0 & 5 & 4 & 250 \\
& FPAFT-df=4 & -0.004 &  0.8 & 94.8 & 3 & 3 & 249 & 0.009 & -1.8 & 92.4 & 4 & 1 & 250 & 0.062 & -12.4 & 85.6 & 6 & 3 & 250 & -0.000 &  0.0 & 94.4 & 6 & 6 & 250 \\
& FPAFT-df=5 & -0.002 &  0.4 & 94.8 & 4 & 4 & 250 & 0.003 & -0.6 & 94.8 & 2 & 3 & 250 & 0.045 & -9.0 & 82.8 & 5 & 2.5 & 250 & 0.000 &  0.0 & 94.8 & 7 & 7 & 249 \\
& FPAFT-df=6 & 0.000 &  0.0 & 91.2 & 5 & 5 & 244 & -0.004 &  0.8 & 94.4 & 3 & 4 & 250 & 0.031 & -6.2 & 86.4 & 3 & 4 & 249 & -0.000 &  0.0 & 89.6 & 8 & 8 & 242 \\
& FPAFT-df=7 & 0.003 & -0.6 & 89.6 & 6 & 6 & 239 & -0.004 &  0.8 & 93.2 & 3 & 5 & 244 & 0.022 & -4.4 & 88.0 & 3 & 6 & 247 & 0.004 & -0.8 & 80.8 & 9 & 9 & 229 \\
& FPAFT-df=8 & 0.007 & -1.4 & 84.8 & 6 & 7 & 237 & 0.001 & -0.2 & 90.4 & 5 & 6 & 244 & 0.016 & -3.2 & 90.0 & 3 & 8 & 248 & -0.000 &  0.0 & 68.4 & 10 & 10 & 216 \\
& FPAFT-df=9 & 0.010 & -2.0 & 78.4 & 5 & 8 & 226 & 0.002 & -0.4 & 83.2 & 6 & 7 & 236 & 0.010 & -2.0 & 88.0 & 4 & 9 & 242 & 0.003 & -0.6 & 60.8 & 11 & 11 & 201 \\
\hline 
\end{tabular}
}
\end{center}
\end{sidewaystable}
\begin{table}[!ht]
\caption{Bias, percentage bias and coverage of estimates of log(-log(S(t))) when X=0 and $\beta=.5$}
\label{tab:simpos0}
\begin{center}
\scalebox{0.65}{
\begin{tabular}{c c r r r r r r r r r r r r r r r r} \hline
\multirow{2}{*}{Time} & \multirow{2}{*}{Model} & \multicolumn{4}{c}{Scenario 1} & \multicolumn{4}{c}{Scenario 2} & \multicolumn{4}{c}{Scenario 3} & \multicolumn{4}{c}{Scenario 4}  \\
 & & Bias & \% Bias & Cov. & \# Conv. & Bias & \% Bias & Cov. & \# Conv.  & Bias & \% Bias & Cov. & \# Conv. & Bias & \% Bias & Cov. & \# Conv.  \\
\hline 
1 & Weibull & 0.223 & -9.7 & 18.4 & 250 & -0.016 &    . & 95.2 & 250 & 0.140 & -13.4 & 34.8 & 250 & -0.003 &  0.1 & 95.2 & 250 \\
2 & Weibull & -0.149 & 38.5 & 17.2 & 250 & -0.131 & -20.6 & 20.4 & 250 & 0.188 & -36.5 &  2.8 & 250 & -0.000 &  0.0 & 95.2 & 250 \\
3 & Weibull & -0.220 & -37.5 &  2.0 & 250 & -0.062 & -7.1 & 75.2 & 250 & 0.108 & -112.6 & 41.2 & 250 & 0.002 & -0.2 & 95.2 & 250 \\
4 & Weibull & 0.085 &  9.2 & 58.8 & 250 & 0.038 &  3.8 & 83.6 & 250 & -0.017 & -6.3 & 97.2 & 250 & 0.003 & -0.5 & 95.2 & 250 \\
5 & Weibull & 0.431 & 40.2 &  0.0 & 250 & 0.121 & 11.3 & 38.8 & 250 & -0.102 & -18.9 & 48.8 & 250 & 0.004 & -1.1 & 95.2 & 250 \\
1 & Gamma & 0.069 & -3.0 & 87.6 & 250 & -0.053 &    . & 81.2 & 250 & 0.011 & -1.1 & 91.6 & 238 & -0.001 &  0.0 & 95.2 & 250 \\
2 & Gamma & -0.023 &  5.9 & 94.8 & 250 & -0.143 & -22.5 & 16.0 & 250 & 0.051 & -9.9 & 78.8 & 238 & -0.001 &  0.1 & 93.2 & 250 \\
3 & Gamma & -0.139 & -23.7 & 16.4 & 250 & -0.047 & -5.4 & 82.0 & 250 & 0.020 & -20.9 & 90.0 & 238 & -0.002 &  0.2 & 92.4 & 250 \\
4 & Gamma & 0.059 &  6.4 & 72.8 & 250 & 0.078 &  7.9 & 66.8 & 250 & -0.024 & -8.9 & 86.8 & 238 & -0.000 &  0.0 & 94.0 & 250 \\
5 & Gamma & 0.285 & 26.6 &  0.4 & 250 & 0.184 & 17.1 & 12.4 & 250 & 0.004 &  0.7 & 90.8 & 238 & 0.004 & -1.1 & 95.6 & 250 \\
1 & GenF & 0.013 & -0.6 & 94.4 & 244 & 0.002 &    . & 74.0 & 191 & 0.015 & -1.4 & 29.2 & 85 & -0.007 &  0.3 & 69.2 & 190 \\
2 & GenF & 0.011 & -2.8 & 94.8 & 247 & -0.017 & -2.7 & 74.4 & 193 & 0.050 & -9.7 & 32.0 & 104 & 0.008 & -0.5 & 70.8 & 196 \\
3 & GenF & -0.034 & -5.8 & 86.4 & 247 & -0.001 & -0.1 & 74.0 & 193 & 0.035 & -36.5 & 60.0 & 181 & 0.016 & -1.6 & 70.8 & 199 \\
4 & GenF & 0.029 &  3.1 & 86.8 & 247 & 0.030 &  3.0 & 68.4 & 193 & 0.037 & 13.7 & 65.2 & 226 & 0.024 & -3.8 & 71.6 & 213 \\
5 & GenF & 0.106 &  9.9 & 46.4 & 247 & 0.046 &  4.3 & 65.2 & 193 & 0.036 &  6.7 & 65.2 & 232 & 0.011 & -3.0 & 72.8 & 216 \\
1 & FPAFT-df=2 & -0.052 &  2.3 & 84.4 & 250 & -0.057 &    . & 76.4 & 250 & 0.065 & -6.2 & 78.4 & 250 & -0.325 & 14.1 & 14.8 & 250 \\
2 & FPAFT-df=2 & -0.044 & 11.4 & 94.0 & 250 & -0.124 & -19.5 & 25.6 & 250 & 0.152 & -29.5 & 18.8 & 250 & -0.307 & 20.9 &  3.6 & 250 \\
3 & FPAFT-df=2 & -0.190 & -32.4 &  2.4 & 250 & -0.021 & -2.4 & 89.2 & 250 & 0.110 & -114.7 & 40.0 & 250 & -0.300 & 30.5 &  0.8 & 250 \\
4 & FPAFT-df=2 & -0.025 & -2.7 & 80.0 & 250 & 0.105 & 10.6 & 51.2 & 250 & 0.016 &  5.9 & 96.4 & 250 & -0.296 & 46.3 &  0.0 & 250 \\
5 & FPAFT-df=2 & 0.190 & 17.7 & 17.2 & 250 & 0.209 & 19.4 &  4.8 & 250 & -0.045 & -8.3 & 88.4 & 250 & -0.293 & 78.9 &  0.0 & 250 \\
1 & FPAFT-df=3 & -0.179 &  7.8 & 59.2 & 250 & -0.000 &    . & 98.0 & 250 & 0.046 & -4.4 & 88.4 & 250 & -0.322 & 14.0 & 20.8 & 250 \\
2 & FPAFT-df=3 & -0.058 & 15.0 & 78.0 & 250 & -0.104 & -16.3 & 39.6 & 250 & 0.076 & -14.8 & 69.6 & 250 & -0.306 & 20.8 &  4.4 & 250 \\
3 & FPAFT-df=3 & -0.115 & -19.6 & 27.6 & 250 & -0.050 & -5.7 & 79.6 & 250 & 0.082 & -85.5 & 61.6 & 250 & -0.301 & 30.6 &  5.6 & 250 \\
4 & FPAFT-df=3 & -0.094 & -10.2 & 44.8 & 250 & 0.035 &  3.5 & 85.2 & 250 & 0.051 & 18.9 & 78.4 & 250 & -0.298 & 46.6 &  0.0 & 250 \\
5 & FPAFT-df=3 & -0.045 & -4.2 & 78.0 & 250 & 0.105 &  9.8 & 52.8 & 250 & 0.047 &  8.7 & 83.2 & 250 & -0.294 & 79.2 &  0.0 & 250 \\
1 & FPAFT-df=4 & -0.105 &  4.6 & 80.8 & 250 & 0.001 &    . & 93.2 & 250 & 0.062 & -5.9 & 75.2 & 250 & -0.319 & 13.9 & 24.0 & 250 \\
2 & FPAFT-df=4 & -0.106 & 27.4 & 59.2 & 250 & -0.036 & -5.7 & 90.4 & 250 & 0.075 & -14.6 & 74.4 & 250 & -0.308 & 20.9 &  6.4 & 250 \\
3 & FPAFT-df=4 & -0.088 & -15.0 & 49.2 & 250 & -0.032 & -3.7 & 87.2 & 250 & 0.073 & -76.1 & 69.6 & 250 & -0.302 & 30.7 &  6.0 & 250 \\
4 & FPAFT-df=4 & -0.087 & -9.4 & 50.8 & 250 & -0.004 & -0.4 & 91.2 & 250 & 0.051 & 18.9 & 78.8 & 250 & -0.297 & 46.5 &  0.8 & 250 \\
5 & FPAFT-df=4 & -0.079 & -7.4 & 68.4 & 250 & 0.017 &  1.6 & 90.8 & 250 & 0.060 & 11.1 & 78.4 & 250 & -0.294 & 79.2 &  0.0 & 250 \\
1 & FPAFT-df=5 & -0.122 &  5.3 & 80.0 & 250 & -0.011 &    . & 93.2 & 250 & 0.062 & -5.9 & 76.4 & 250 & -0.318 & 13.8 & 24.4 & 250 \\
2 & FPAFT-df=5 & -0.106 & 27.4 & 59.6 & 250 & -0.025 & -3.9 & 93.6 & 250 & 0.075 & -14.6 & 72.8 & 250 & -0.309 & 21.0 &  9.2 & 250 \\
3 & FPAFT-df=5 & -0.082 & -14.0 & 53.6 & 250 & -0.025 & -2.9 & 92.0 & 250 & 0.073 & -76.1 & 75.2 & 250 & -0.299 & 30.4 &  7.2 & 250 \\
4 & FPAFT-df=5 & -0.080 & -8.7 & 56.0 & 250 & -0.007 & -0.7 & 91.2 & 250 & 0.051 & 18.9 & 80.4 & 250 & -0.298 & 46.6 &  1.6 & 250 \\
5 & FPAFT-df=5 & -0.088 & -8.2 & 65.2 & 250 & 0.005 &  0.5 & 92.0 & 250 & 0.058 & 10.7 & 79.6 & 250 & -0.294 & 79.2 &  0.4 & 250 \\
1 & FPAFT-df=6 & -0.118 &  5.1 & 81.2 & 250 & -0.022 &    . & 90.4 & 250 & 0.063 & -6.0 & 76.8 & 250 & -0.318 & 13.8 & 24.8 & 250 \\
2 & FPAFT-df=6 & -0.099 & 25.6 & 63.6 & 250 & -0.018 & -2.8 & 94.4 & 250 & 0.072 & -14.0 & 74.0 & 250 & -0.308 & 20.9 &  8.4 & 250 \\
3 & FPAFT-df=6 & -0.080 & -13.6 & 55.2 & 250 & -0.019 & -2.2 & 92.8 & 250 & 0.076 & -79.2 & 72.0 & 250 & -0.299 & 30.4 &  9.2 & 250 \\
4 & FPAFT-df=6 & -0.076 & -8.3 & 57.2 & 250 & -0.008 & -0.8 & 92.8 & 250 & 0.052 & 19.3 & 80.0 & 250 & -0.297 & 46.5 &  1.6 & 250 \\
5 & FPAFT-df=6 & -0.090 & -8.4 & 65.2 & 250 & -0.004 & -0.4 & 93.2 & 250 & 0.056 & 10.4 & 79.6 & 250 & -0.294 & 79.2 &  0.0 & 250 \\
1 & FPAFT-df=7 & -0.119 &  5.2 & 80.4 & 249 & -0.023 &    . & 90.4 & 250 & 0.062 & -5.9 & 77.2 & 250 & -0.318 & 13.8 & 24.4 & 250 \\
2 & FPAFT-df=7 & -0.096 & 24.8 & 65.2 & 249 & -0.015 & -2.4 & 94.4 & 250 & 0.072 & -14.0 & 74.4 & 250 & -0.303 & 20.6 & 13.2 & 250 \\
3 & FPAFT-df=7 & -0.081 & -13.8 & 54.0 & 249 & -0.014 & -1.6 & 94.8 & 250 & 0.078 & -81.3 & 69.6 & 250 & -0.298 & 30.3 & 10.0 & 250 \\
4 & FPAFT-df=7 & -0.071 & -7.7 & 62.4 & 249 & -0.007 & -0.7 & 94.0 & 250 & 0.054 & 20.0 & 80.8 & 250 & -0.295 & 46.2 &  3.2 & 250 \\
5 & FPAFT-df=7 & -0.089 & -8.3 & 64.0 & 249 & -0.006 & -0.6 & 93.2 & 250 & 0.056 & 10.4 & 81.2 & 250 & -0.294 & 79.2 &  1.2 & 250 \\
1 & FPAFT-df=8 & -0.117 &  5.1 & 80.8 & 249 & -0.020 &    . & 92.8 & 250 & 0.063 & -6.0 & 77.2 & 250 & -0.317 & 13.8 & 27.2 & 250 \\
2 & FPAFT-df=8 & -0.097 & 25.0 & 66.4 & 249 & -0.014 & -2.2 & 94.0 & 250 & 0.073 & -14.2 & 73.2 & 250 & -0.306 & 20.8 & 10.0 & 250 \\
3 & FPAFT-df=8 & -0.081 & -13.8 & 54.8 & 249 & -0.012 & -1.4 & 94.0 & 250 & 0.078 & -81.3 & 69.6 & 250 & -0.298 & 30.3 & 10.4 & 250 \\
4 & FPAFT-df=8 & -0.069 & -7.5 & 63.2 & 249 & -0.007 & -0.7 & 93.6 & 250 & 0.056 & 20.8 & 78.4 & 250 & -0.295 & 46.2 &  2.8 & 250 \\
5 & FPAFT-df=8 & -0.089 & -8.3 & 66.4 & 249 & -0.009 & -0.8 & 92.0 & 250 & 0.055 & 10.2 & 81.2 & 250 & -0.294 & 79.2 &  0.8 & 250 \\
1 & FPAFT-df=9 & -0.117 &  5.1 & 79.2 & 250 & -0.019 &    . & 90.4 & 250 & 0.063 & -6.0 & 77.2 & 250 & -0.315 & 13.7 & 28.0 & 250 \\
2 & FPAFT-df=9 & -0.098 & 25.3 & 66.0 & 250 & -0.015 & -2.4 & 93.6 & 250 & 0.075 & -14.6 & 73.2 & 250 & -0.304 & 20.7 &  9.6 & 250 \\
3 & FPAFT-df=9 & -0.083 & -14.2 & 54.0 & 250 & -0.010 & -1.1 & 94.4 & 250 & 0.077 & -80.3 & 70.8 & 250 & -0.300 & 30.5 &  7.2 & 250 \\
4 & FPAFT-df=9 & -0.068 & -7.4 & 64.8 & 250 & -0.007 & -0.7 & 93.6 & 250 & 0.057 & 21.2 & 77.6 & 250 & -0.297 & 46.5 &  1.6 & 250 \\
5 & FPAFT-df=9 & -0.089 & -8.3 & 64.8 & 250 & -0.010 & -0.9 & 92.4 & 250 & 0.055 & 10.2 & 81.6 & 250 & -0.293 & 78.9 &  0.8 & 250 \\
\hline 
\end{tabular}
}
\end{center}
\end{table}
\begin{table}[!ht]
\caption{Bias, percentage bias and coverage of estimates of log(-log(S(t))) when X=1 and $\beta=.5$}
\label{tab:simpos1}
\begin{center}
\scalebox{0.65}{
\begin{tabular}{c c r r r r r r r r r r r r r r r r} \hline
\multirow{2}{*}{Time} & \multirow{2}{*}{Model} & \multicolumn{4}{c}{Scenario 1} & \multicolumn{4}{c}{Scenario 2} & \multicolumn{4}{c}{Scenario 3} & \multicolumn{4}{c}{Scenario 4}  \\
 & & Bias & \% Bias & Cov. & \# Conv. & Bias & \% Bias & Cov. & \# Conv.  & Bias & \% Bias & Cov. & \# Conv. & Bias & \% Bias & Cov. & \# Conv.  \\
\hline 
1 & Weibull & 0.610 & -16.9 &  0.0 & 250 & 0.132 & -27.1 & 24.0 & 250 & -0.050 &  3.6 & 88.8 & 250 & -0.008 &  0.3 & 96.8 & 250 \\
2 & Weibull & 0.306 & -17.3 &  0.0 & 250 & -0.026 & -13.5 & 92.4 & 250 & 0.050 & -5.5 & 87.2 & 250 & -0.005 &  0.2 & 96.8 & 250 \\
3 & Weibull & 0.082 & -12.7 & 64.8 & 250 & -0.091 & -16.2 & 51.6 & 250 & 0.076 & -12.7 & 74.8 & 250 & -0.003 &  0.2 & 97.2 & 250 \\
4 & Weibull & -0.043 & -36.1 & 88.4 & 250 & -0.076 & -9.9 & 68.8 & 250 & 0.050 & -15.1 & 87.2 & 250 & -0.002 &  0.2 & 97.2 & 250 \\
5 & Weibull & -0.032 & -5.3 & 92.0 & 250 & -0.022 & -2.5 & 94.0 & 250 & -0.011 & 13.3 & 96.0 & 250 & -0.001 &  0.1 & 96.8 & 250 \\
1 & Gamma & -0.121 &  3.3 & 84.4 & 250 & 0.110 & -22.5 & 37.2 & 250 & -0.034 &  2.4 & 89.6 & 238 & -0.007 &  0.2 & 96.8 & 250 \\
2 & Gamma & 0.186 & -10.5 & 31.2 & 250 & -0.038 & -19.8 & 88.0 & 250 & 0.013 & -1.4 & 92.4 & 238 & -0.002 &  0.1 & 97.2 & 250 \\
3 & Gamma & 0.092 & -14.2 & 61.2 & 250 & -0.087 & -15.4 & 55.2 & 250 & 0.035 & -5.9 & 86.0 & 238 & -0.002 &  0.1 & 96.4 & 250 \\
4 & Gamma & -0.036 & -30.2 & 89.2 & 250 & -0.053 & -6.9 & 80.8 & 250 & 0.029 & -8.8 & 89.6 & 238 & -0.002 &  0.2 & 96.8 & 250 \\
5 & Gamma & -0.075 & -12.4 & 67.6 & 250 & 0.018 &  2.1 & 94.0 & 250 & 0.001 & -1.2 & 92.0 & 238 & -0.001 &  0.1 & 97.2 & 250 \\
1 & GenF & 0.056 & -1.5 & 91.6 & 243 & 0.037 & -7.6 & 67.2 & 183 & -0.035 &  2.5 & 24.4 & 67 & -0.036 &  1.2 & 70.8 & 187 \\
2 & GenF & -0.012 &  0.7 & 93.2 & 245 & 0.001 &  0.5 & 75.2 & 193 & 0.013 & -1.4 & 30.0 & 89 & -0.021 &  1.0 & 72.0 & 191 \\
3 & GenF & -0.005 &  0.8 & 92.8 & 247 & -0.014 & -2.5 & 72.4 & 193 & 0.038 & -6.4 & 36.4 & 101 & -0.013 &  0.8 & 72.0 & 194 \\
4 & GenF & 0.009 &  7.6 & 96.8 & 247 & -0.017 & -2.2 & 70.0 & 193 & 0.033 & -10.0 & 39.2 & 107 & -0.006 &  0.5 & 72.8 & 195 \\
5 & GenF & -0.037 & -6.1 & 85.2 & 247 & -0.000 &  0.0 & 72.8 & 193 & 0.003 & -3.6 & 64.8 & 176 & 0.001 & -0.1 & 75.2 & 200 \\
1 & FPAFT-df=2 & -0.431 & 11.9 & 43.6 & 250 & 0.093 & -19.1 & 52.0 & 250 & -0.099 &  7.1 & 68.0 & 250 & -0.348 & 12.0 & 27.6 & 250 \\
2 & FPAFT-df=2 & 0.048 & -2.7 & 80.0 & 250 & -0.031 & -16.1 & 90.0 & 250 & 0.003 & -0.3 & 96.8 & 250 & -0.326 & 15.7 & 16.0 & 250 \\
3 & FPAFT-df=2 & 0.018 & -2.8 & 92.0 & 250 & -0.067 & -11.9 & 70.8 & 250 & 0.050 & -8.4 & 88.0 & 250 & -0.315 & 19.9 & 14.0 & 250 \\
4 & FPAFT-df=2 & -0.109 & -91.5 & 44.0 & 250 & -0.027 & -3.5 & 91.6 & 250 & 0.047 & -14.2 & 88.4 & 250 & -0.309 & 24.9 & 10.8 & 250 \\
5 & FPAFT-df=2 & -0.168 & -27.7 & 12.8 & 250 & 0.047 &  5.4 & 82.8 & 250 & 0.007 & -8.4 & 98.0 & 250 & -0.305 & 31.4 &  8.8 & 250 \\
1 & FPAFT-df=3 & -0.191 &  5.3 & 84.0 & 250 & 0.050 & -10.2 & 84.4 & 250 & 0.036 & -2.6 & 94.4 & 250 & -0.348 & 12.0 & 36.4 & 250 \\
2 & FPAFT-df=3 & -0.230 & 13.0 & 25.6 & 250 & -0.053 & -27.6 & 79.6 & 250 & 0.031 & -3.4 & 95.6 & 250 & -0.325 & 15.7 & 20.4 & 250 \\
3 & FPAFT-df=3 & -0.149 & 23.1 & 41.2 & 250 & -0.113 & -20.1 & 30.4 & 250 & 0.048 & -8.0 & 87.2 & 250 & -0.317 & 20.0 & 13.6 & 250 \\
4 & FPAFT-df=3 & -0.100 & -84.0 & 51.2 & 250 & -0.105 & -13.7 & 40.8 & 250 & 0.061 & -18.5 & 80.0 & 250 & -0.312 & 25.2 & 13.2 & 250 \\
5 & FPAFT-df=3 & -0.149 & -24.6 &  8.8 & 250 & -0.059 & -6.7 & 74.8 & 250 & 0.056 & -67.5 & 83.2 & 250 & -0.306 & 31.5 & 13.2 & 250 \\
1 & FPAFT-df=4 & -0.135 &  3.7 & 88.8 & 250 & -0.058 & 11.9 & 80.0 & 250 & 0.054 & -3.9 & 87.2 & 250 & -0.343 & 11.8 & 38.0 & 250 \\
2 & FPAFT-df=4 & -0.145 &  8.2 & 62.8 & 250 & -0.018 & -9.4 & 89.6 & 250 & 0.064 & -7.0 & 75.2 & 250 & -0.321 & 15.5 & 18.8 & 250 \\
3 & FPAFT-df=4 & -0.152 & 23.5 & 34.4 & 250 & -0.045 & -8.0 & 80.0 & 250 & 0.071 & -11.9 & 69.2 & 250 & -0.316 & 19.9 & 14.8 & 250 \\
4 & FPAFT-df=4 & -0.093 & -78.1 & 60.0 & 250 & -0.055 & -7.2 & 69.2 & 250 & 0.075 & -22.7 & 71.6 & 250 & -0.313 & 25.3 & 14.4 & 250 \\
5 & FPAFT-df=4 & -0.108 & -17.8 & 29.6 & 250 & -0.042 & -4.8 & 76.4 & 250 & 0.071 & -85.6 & 79.6 & 250 & -0.307 & 31.6 & 13.2 & 250 \\
1 & FPAFT-df=5 & -0.140 &  3.9 & 90.0 & 250 & -0.050 & 10.2 & 82.8 & 250 & 0.055 & -4.0 & 86.0 & 250 & -0.342 & 11.8 & 43.6 & 250 \\
2 & FPAFT-df=5 & -0.141 &  8.0 & 64.0 & 250 & -0.023 & -12.0 & 88.4 & 250 & 0.059 & -6.5 & 80.8 & 250 & -0.322 & 15.5 & 22.8 & 250 \\
3 & FPAFT-df=5 & -0.124 & 19.2 & 48.8 & 250 & -0.033 & -5.9 & 85.6 & 250 & 0.066 & -11.1 & 71.6 & 250 & -0.320 & 20.2 & 18.0 & 250 \\
4 & FPAFT-df=5 & -0.102 & -85.6 & 55.6 & 250 & -0.043 & -5.6 & 79.6 & 250 & 0.072 & -21.8 & 72.0 & 250 & -0.315 & 25.4 & 13.2 & 250 \\
5 & FPAFT-df=5 & -0.096 & -15.9 & 45.2 & 250 & -0.034 & -3.9 & 82.4 & 250 & 0.068 & -81.9 & 80.8 & 250 & -0.306 & 31.5 & 15.2 & 250 \\
1 & FPAFT-df=6 & -0.131 &  3.6 & 90.4 & 250 & -0.030 &  6.1 & 88.8 & 250 & 0.052 & -3.7 & 87.2 & 250 & -0.344 & 11.9 & 43.2 & 250 \\
2 & FPAFT-df=6 & -0.133 &  7.5 & 70.4 & 250 & -0.029 & -15.1 & 86.4 & 250 & 0.060 & -6.6 & 81.6 & 250 & -0.328 & 15.8 & 22.8 & 250 \\
3 & FPAFT-df=6 & -0.116 & 18.0 & 58.8 & 250 & -0.025 & -4.4 & 86.8 & 250 & 0.063 & -10.6 & 79.2 & 250 & -0.324 & 20.5 & 17.6 & 250 \\
4 & FPAFT-df=6 & -0.105 & -88.2 & 54.8 & 250 & -0.031 & -4.1 & 85.2 & 250 & 0.070 & -21.2 & 73.6 & 250 & -0.316 & 25.5 & 12.8 & 250 \\
5 & FPAFT-df=6 & -0.091 & -15.0 & 48.8 & 250 & -0.025 & -2.8 & 86.4 & 250 & 0.068 & -81.9 & 78.4 & 250 & -0.311 & 32.0 & 14.0 & 250 \\
1 & FPAFT-df=7 & -0.130 &  3.6 & 89.6 & 249 & -0.027 &  5.5 & 88.0 & 250 & 0.052 & -3.7 & 87.6 & 250 & -0.344 & 11.9 & 43.6 & 250 \\
2 & FPAFT-df=7 & -0.126 &  7.1 & 69.2 & 249 & -0.031 & -16.1 & 86.4 & 250 & 0.060 & -6.6 & 80.8 & 250 & -0.332 & 16.0 & 24.8 & 250 \\
3 & FPAFT-df=7 & -0.113 & 17.5 & 58.8 & 249 & -0.020 & -3.6 & 87.2 & 250 & 0.065 & -10.9 & 73.6 & 250 & -0.324 & 20.5 & 18.4 & 250 \\
4 & FPAFT-df=7 & -0.103 & -86.5 & 51.2 & 249 & -0.023 & -3.0 & 86.4 & 250 & 0.070 & -21.2 & 73.6 & 250 & -0.317 & 25.6 & 16.0 & 250 \\
5 & FPAFT-df=7 & -0.087 & -14.4 & 50.8 & 249 & -0.018 & -2.1 & 86.4 & 250 & 0.070 & -84.3 & 76.8 & 250 & -0.312 & 32.1 & 16.8 & 250 \\
1 & FPAFT-df=8 & -0.132 &  3.6 & 89.6 & 249 & -0.028 &  5.7 & 86.0 & 250 & 0.052 & -3.7 & 87.6 & 250 & -0.340 & 11.7 & 44.4 & 250 \\
2 & FPAFT-df=8 & -0.124 &  7.0 & 72.4 & 249 & -0.031 & -16.1 & 81.6 & 250 & 0.059 & -6.5 & 80.4 & 250 & -0.325 & 15.7 & 29.6 & 250 \\
3 & FPAFT-df=8 & -0.112 & 17.3 & 63.2 & 249 & -0.021 & -3.7 & 84.0 & 250 & 0.065 & -10.9 & 74.0 & 250 & -0.322 & 20.3 & 17.6 & 250 \\
4 & FPAFT-df=8 & -0.102 & -85.6 & 53.2 & 249 & -0.021 & -2.7 & 84.4 & 250 & 0.068 & -20.6 & 74.0 & 250 & -0.313 & 25.3 & 16.0 & 250 \\
5 & FPAFT-df=8 & -0.088 & -14.5 & 51.2 & 249 & -0.016 & -1.8 & 85.6 & 250 & 0.070 & -84.3 & 74.8 & 250 & -0.308 & 31.7 & 18.8 & 250 \\
1 & FPAFT-df=9 & -0.133 &  3.7 & 92.0 & 250 & -0.025 &  5.1 & 84.0 & 250 & 0.051 & -3.7 & 86.4 & 250 & -0.342 & 11.8 & 46.0 & 250 \\
2 & FPAFT-df=9 & -0.121 &  6.8 & 72.8 & 250 & -0.028 & -14.6 & 81.2 & 250 & 0.059 & -6.5 & 80.4 & 250 & -0.328 & 15.8 & 26.8 & 250 \\
3 & FPAFT-df=9 & -0.111 & 17.2 & 59.6 & 250 & -0.020 & -3.6 & 82.8 & 250 & 0.065 & -10.9 & 74.8 & 250 & -0.322 & 20.3 & 16.8 & 250 \\
4 & FPAFT-df=9 & -0.100 & -84.0 & 53.2 & 250 & -0.018 & -2.4 & 85.6 & 250 & 0.067 & -20.3 & 71.2 & 250 & -0.315 & 25.4 & 14.8 & 250 \\
5 & FPAFT-df=9 & -0.088 & -14.5 & 51.6 & 250 & -0.013 & -1.5 & 86.8 & 250 & 0.069 & -83.1 & 73.2 & 250 & -0.310 & 31.9 & 13.2 & 250 \\
\hline 
\end{tabular}
}
\end{center}
\end{table}
\begin{table}[!ht]
\caption{Bias, percentage bias and coverage of estimates of log(-log(S(t))) when X=0 and $\beta=-.5$}
\label{tab:simneg0}
\begin{center}
\scalebox{0.65}{
\begin{tabular}{c c r r r r r r r r r r r r r r r r} \hline
\multirow{2}{*}{Time} & \multirow{2}{*}{Model} & \multicolumn{4}{c}{Scenario 1} & \multicolumn{4}{c}{Scenario 2} & \multicolumn{4}{c}{Scenario 3} & \multicolumn{4}{c}{Scenario 4}  \\
 & & Bias & \% Bias & Cov. & \# Conv. & Bias & \% Bias & Cov. & \# Conv.  & Bias & \% Bias & Cov. & \# Conv. & Bias & \% Bias & Cov. & \# Conv.  \\
\hline 
1 & Weibull & 0.590 & -25.6 &  0.0 & 250 & -0.003 &    . & 96.0 & 250 & 0.096 & -9.2 & 60.8 & 250 & 0.000 &  0.0 & 96.4 & 250 \\
2 & Weibull & -0.014 &  3.6 & 94.4 & 250 & -0.139 & -21.8 & 15.2 & 250 & 0.170 & -33.0 &  9.6 & 250 & 0.002 & -0.1 & 97.2 & 250 \\
3 & Weibull & -0.221 & -37.7 &  0.4 & 250 & -0.083 & -9.5 & 55.6 & 250 & 0.105 & -109.5 & 41.6 & 250 & 0.003 & -0.3 & 95.2 & 250 \\
4 & Weibull & -0.011 & -1.2 & 89.6 & 250 & 0.009 &  0.9 & 93.6 & 250 & -0.008 & -3.0 & 97.6 & 250 & 0.003 & -0.5 & 96.0 & 250 \\
5 & Weibull & 0.259 & 24.1 &  0.4 & 250 & 0.085 &  7.9 & 61.2 & 250 & -0.085 & -15.7 & 59.2 & 250 & 0.004 & -1.1 & 94.8 & 250 \\
1 & Gamma & 0.192 & -8.3 & 46.4 & 250 & -0.010 &    . & 95.2 & 250 & 0.067 & -6.4 & 72.8 & 250 & -0.002 &  0.1 & 95.6 & 250 \\
2 & Gamma & 0.034 & -8.8 & 92.0 & 250 & -0.140 & -22.0 & 14.8 & 250 & 0.121 & -23.5 & 26.0 & 250 & 0.003 & -0.2 & 97.2 & 250 \\
3 & Gamma & -0.166 & -28.3 &  4.0 & 250 & -0.078 & -8.9 & 62.0 & 250 & 0.079 & -82.4 & 60.8 & 250 & 0.003 & -0.3 & 95.6 & 250 \\
4 & Gamma & -0.036 & -3.9 & 86.0 & 250 & 0.019 &  1.9 & 91.6 & 250 & 0.002 &  0.7 & 97.2 & 250 & 0.003 & -0.5 & 96.0 & 250 \\
5 & Gamma & 0.135 & 12.6 & 19.6 & 250 & 0.101 &  9.4 & 54.0 & 250 & -0.031 & -5.7 & 92.0 & 250 & 0.003 & -0.8 & 94.8 & 250 \\
1 & GenF & 0.013 & -0.6 & 94.8 & 244 & 0.004 &    . & 70.8 & 183 & -0.007 &  0.7 & 18.0 & 45 & -0.013 &  0.6 & 86.0 & 233 \\
2 & GenF & 0.011 & -2.8 & 94.0 & 248 & -0.023 & -3.6 & 69.6 & 185 & 0.047 & -9.1 & 30.0 & 87 & -0.011 &  0.7 & 86.8 & 234 \\
3 & GenF & -0.049 & -8.4 & 79.6 & 248 & -0.010 & -1.1 & 71.2 & 185 & 0.025 & -26.1 & 38.4 & 98 & -0.006 &  0.6 & 85.2 & 234 \\
4 & GenF & 0.001 &  0.1 & 93.2 & 248 & 0.019 &  1.9 & 66.8 & 185 & -0.004 & -1.5 & 42.4 & 109 & -0.001 &  0.2 & 85.2 & 234 \\
5 & GenF & 0.071 &  6.6 & 60.8 & 248 & 0.034 &  3.2 & 65.6 & 185 & -0.017 & -3.1 & 41.6 & 109 & 0.000 &  0.0 & 84.8 & 238 \\
1 & FPAFT-df=2 & 0.111 & -4.8 & 72.4 & 250 & -0.033 &    . & 88.8 & 250 & 0.012 & -1.1 & 94.4 & 250 & 0.309 & -13.4 &  3.2 & 250 \\
2 & FPAFT-df=2 & 0.146 & -37.7 & 21.2 & 250 & -0.144 & -22.6 & 14.0 & 250 & 0.103 & -20.0 & 45.6 & 250 & 0.305 & -20.7 &  0.4 & 250 \\
3 & FPAFT-df=2 & -0.079 & -13.5 & 53.2 & 250 & -0.067 & -7.7 & 69.6 & 250 & 0.071 & -74.0 & 68.0 & 250 & 0.301 & -30.6 &  0.4 & 250 \\
4 & FPAFT-df=2 & -0.018 & -2.0 & 94.4 & 250 & 0.039 &  3.9 & 85.6 & 250 & -0.013 & -4.8 & 95.6 & 250 & 0.298 & -46.6 &  0.0 & 250 \\
5 & FPAFT-df=2 & 0.084 &  7.8 & 57.2 & 250 & 0.128 & 11.9 & 36.8 & 250 & -0.063 & -11.7 & 76.4 & 250 & 0.295 & -79.5 &  0.0 & 250 \\
1 & FPAFT-df=3 & 0.076 & -3.3 & 88.8 & 250 & 0.057 &    . & 79.2 & 250 & 0.040 & -3.8 & 89.2 & 250 & 0.309 & -13.4 &  5.2 & 250 \\
2 & FPAFT-df=3 & 0.101 & -26.1 & 49.6 & 250 & -0.078 & -12.3 & 61.2 & 250 & 0.059 & -11.5 & 78.4 & 250 & 0.307 & -20.9 &  0.4 & 250 \\
3 & FPAFT-df=3 & 0.004 &  0.7 & 91.6 & 250 & -0.061 & -7.0 & 73.2 & 250 & 0.018 & -18.8 & 93.6 & 250 & 0.302 & -30.7 &  0.4 & 250 \\
4 & FPAFT-df=3 & 0.039 &  4.2 & 84.8 & 250 & -0.009 & -0.9 & 93.2 & 250 & -0.045 & -16.7 & 82.8 & 250 & 0.298 & -46.6 &  0.0 & 250 \\
5 & FPAFT-df=3 & 0.076 &  7.1 & 57.6 & 250 & 0.032 &  3.0 & 87.6 & 250 & -0.065 & -12.0 & 74.8 & 250 & 0.294 & -79.2 &  0.0 & 250 \\
1 & FPAFT-df=4 & 0.081 & -3.5 & 84.4 & 249 & 0.038 &    . & 81.6 & 250 & -0.021 &  2.0 & 93.6 & 250 & 0.310 & -13.5 &  5.6 & 250 \\
2 & FPAFT-df=4 & 0.095 & -24.5 & 60.8 & 249 & 0.001 &  0.2 & 94.8 & 250 & 0.026 & -5.1 & 89.6 & 250 & 0.304 & -20.7 &  0.8 & 250 \\
3 & FPAFT-df=4 & 0.005 &  0.9 & 90.4 & 249 & 0.000 &  0.0 & 95.6 & 250 & 0.044 & -45.9 & 82.4 & 250 & 0.304 & -30.9 &  0.4 & 250 \\
4 & FPAFT-df=4 & 0.045 &  4.9 & 81.2 & 249 & 0.016 &  1.6 & 93.6 & 250 & -0.008 & -3.0 & 96.0 & 250 & 0.300 & -46.9 &  0.4 & 250 \\
5 & FPAFT-df=4 & 0.081 &  7.6 & 55.2 & 249 & 0.020 &  1.9 & 93.6 & 250 & -0.045 & -8.3 & 87.2 & 250 & 0.294 & -79.2 &  0.0 & 250 \\
1 & FPAFT-df=5 & 0.083 & -3.6 & 85.6 & 250 & 0.031 &    . & 88.0 & 250 & 0.007 & -0.7 & 95.6 & 250 & 0.310 & -13.5 &  5.2 & 250 \\
2 & FPAFT-df=5 & 0.098 & -25.3 & 59.2 & 250 & 0.006 &  0.9 & 92.0 & 250 & -0.003 &  0.6 & 95.6 & 250 & 0.304 & -20.7 &  0.8 & 250 \\
3 & FPAFT-df=5 & 0.004 &  0.7 & 89.2 & 250 & 0.007 &  0.8 & 95.2 & 250 & 0.028 & -29.2 & 85.2 & 250 & 0.304 & -30.9 &  0.4 & 250 \\
4 & FPAFT-df=5 & 0.042 &  4.6 & 82.0 & 250 & 0.021 &  2.1 & 92.0 & 250 & 0.015 &  5.6 & 91.2 & 250 & 0.300 & -46.9 &  0.0 & 250 \\
5 & FPAFT-df=5 & 0.078 &  7.3 & 57.6 & 250 & 0.022 &  2.0 & 93.2 & 250 & -0.023 & -4.3 & 91.6 & 250 & 0.294 & -79.2 &  0.0 & 250 \\
1 & FPAFT-df=6 & 0.076 & -3.3 & 86.8 & 250 & 0.025 &    . & 90.8 & 250 & 0.004 & -0.4 & 96.4 & 250 & 0.311 & -13.5 &  6.0 & 250 \\
2 & FPAFT-df=6 & 0.089 & -23.0 & 64.8 & 250 & 0.008 &  1.3 & 90.8 & 250 & 0.005 & -1.0 & 96.4 & 250 & 0.305 & -20.7 &  0.4 & 250 \\
3 & FPAFT-df=6 & 0.005 &  0.9 & 90.8 & 250 & 0.013 &  1.5 & 93.6 & 250 & 0.010 & -10.4 & 88.0 & 250 & 0.304 & -30.9 &  0.4 & 250 \\
4 & FPAFT-df=6 & 0.036 &  3.9 & 85.2 & 250 & 0.025 &  2.5 & 90.4 & 250 & 0.020 &  7.4 & 89.2 & 250 & 0.301 & -47.1 &  0.0 & 250 \\
5 & FPAFT-df=6 & 0.073 &  6.8 & 61.6 & 250 & 0.024 &  2.2 & 93.2 & 250 & -0.010 & -1.8 & 93.6 & 250 & 0.294 & -79.2 &  0.0 & 250 \\
1 & FPAFT-df=7 & 0.076 & -3.3 & 86.0 & 248 & 0.027 &    . & 91.6 & 250 & 0.002 & -0.2 & 96.4 & 249 & 0.310 & -13.5 &  6.4 & 250 \\
2 & FPAFT-df=7 & 0.084 & -21.7 & 67.2 & 248 & 0.007 &  1.1 & 90.4 & 250 & 0.010 & -1.9 & 94.8 & 249 & 0.307 & -20.9 &  1.6 & 250 \\
3 & FPAFT-df=7 & 0.009 &  1.5 & 90.4 & 248 & 0.015 &  1.7 & 93.6 & 250 & 0.005 & -5.2 & 92.0 & 249 & 0.305 & -31.0 &  0.8 & 250 \\
4 & FPAFT-df=7 & 0.031 &  3.4 & 88.0 & 248 & 0.028 &  2.8 & 91.2 & 250 & 0.018 &  6.7 & 87.2 & 249 & 0.301 & -47.1 &  0.8 & 250 \\
5 & FPAFT-df=7 & 0.068 &  6.3 & 66.4 & 248 & 0.026 &  2.4 & 92.0 & 250 & -0.003 & -0.6 & 93.6 & 249 & 0.295 & -79.5 &  0.8 & 250 \\
1 & FPAFT-df=8 & 0.075 & -3.3 & 85.6 & 249 & 0.026 &    . & 89.6 & 250 & 0.004 & -0.4 & 95.6 & 250 & 0.307 & -13.3 &  8.8 & 249 \\
2 & FPAFT-df=8 & 0.085 & -21.9 & 65.6 & 249 & 0.007 &  1.1 & 90.8 & 250 & 0.008 & -1.6 & 95.6 & 250 & 0.306 & -20.8 &  1.2 & 249 \\
3 & FPAFT-df=8 & 0.015 &  2.6 & 89.6 & 249 & 0.014 &  1.6 & 92.4 & 250 & 0.006 & -6.3 & 91.6 & 250 & 0.304 & -30.9 &  1.2 & 249 \\
4 & FPAFT-df=8 & 0.026 &  2.8 & 89.2 & 249 & 0.027 &  2.7 & 90.4 & 250 & 0.014 &  5.2 & 84.8 & 250 & 0.299 & -46.8 &  0.4 & 249 \\
5 & FPAFT-df=8 & 0.064 &  6.0 & 68.8 & 249 & 0.025 &  2.3 & 91.2 & 250 & 0.000 &  0.0 & 92.0 & 250 & 0.292 & -78.7 &  1.2 & 249 \\
1 & FPAFT-df=9 & 0.075 & -3.3 & 85.2 & 249 & 0.024 &    . & 91.2 & 250 & 0.004 & -0.4 & 95.2 & 249 & 0.305 & -13.2 & 12.8 & 249 \\
2 & FPAFT-df=9 & 0.085 & -21.9 & 64.4 & 249 & 0.007 &  1.1 & 90.8 & 250 & 0.007 & -1.4 & 95.6 & 249 & 0.304 & -20.7 &  2.4 & 249 \\
3 & FPAFT-df=9 & 0.023 &  3.9 & 89.2 & 249 & 0.013 &  1.5 & 92.4 & 250 & 0.008 & -8.3 & 93.2 & 249 & 0.302 & -30.7 &  0.8 & 249 \\
4 & FPAFT-df=9 & 0.022 &  2.4 & 90.0 & 249 & 0.026 &  2.6 & 90.8 & 250 & 0.010 &  3.7 & 85.2 & 249 & 0.299 & -46.8 &  2.4 & 249 \\
5 & FPAFT-df=9 & 0.059 &  5.5 & 74.0 & 249 & 0.025 &  2.3 & 91.6 & 250 & 0.002 &  0.4 & 92.0 & 249 & 0.293 & -78.9 &  1.6 & 249 \\
\hline 
\end{tabular}
}
\end{center}
\end{table}
\begin{table}[!ht]
\caption{Bias, percentage bias and coverage of estimates of log(-log(S(t))) when X=1 and $\beta=-.5$}
\label{tab:simneg1}
\begin{center}
\scalebox{0.65}{
\begin{tabular}{c c r r r r r r r r r r r r r r r r} \hline
\multirow{2}{*}{Time} & \multirow{2}{*}{Model} & \multicolumn{4}{c}{Scenario 1} & \multicolumn{4}{c}{Scenario 2} & \multicolumn{4}{c}{Scenario 3} & \multicolumn{4}{c}{Scenario 4}  \\
 & & Bias & \% Bias & Cov. & \# Conv. & Bias & \% Bias & Cov. & \# Conv.  & Bias & \% Bias & Cov. & \# Conv. & Bias & \% Bias & Cov. & \# Conv.  \\
\hline 
1 & Weibull & 0.046 & -5.0 & 87.2 & 250 & -0.152 & -31.7 & 10.4 & 250 & 0.157 & -23.2 & 15.6 & 250 & -0.003 &  0.2 & 96.0 & 250 \\
2 & Weibull & -0.298 & -40.5 &  0.0 & 250 & -0.085 & -9.3 & 57.2 & 250 & 0.066 & 334.4 & 73.6 & 250 & -0.001 &  0.1 & 96.0 & 250 \\
3 & Weibull & 0.140 & 13.1 & 28.8 & 250 & 0.051 &  4.8 & 78.8 & 250 & -0.089 & -16.8 & 52.0 & 250 & -0.000 &  0.0 & 96.4 & 250 \\
4 & Weibull & 0.453 & 35.0 &  0.0 & 250 & 0.148 & 12.5 & 24.8 & 250 & -0.082 & -10.6 & 61.6 & 250 & 0.000 &  0.0 & 96.8 & 250 \\
5 & Weibull & 0.660 & 43.7 &  0.0 & 250 & 0.220 & 17.3 &  2.8 & 250 & -0.027 & -3.0 & 90.4 & 250 & 0.001 &  0.4 & 95.2 & 250 \\
1 & Gamma & 0.115 & -12.5 & 54.0 & 250 & -0.160 & -33.3 &  6.4 & 250 & 0.049 & -7.2 & 86.4 & 250 & 0.000 &  0.0 & 95.2 & 250 \\
2 & Gamma & -0.173 & -23.5 &  2.8 & 250 & -0.083 & -9.1 & 59.2 & 250 & -0.020 & -101.3 & 92.8 & 250 & 0.001 & -0.1 & 96.8 & 250 \\
3 & Gamma & 0.113 & 10.6 & 35.6 & 250 & 0.062 &  5.8 & 74.4 & 250 & -0.115 & -21.7 & 33.2 & 250 & 0.001 & -0.3 & 96.0 & 250 \\
4 & Gamma & 0.260 & 20.1 &  0.0 & 250 & 0.166 & 14.0 & 21.2 & 250 & -0.037 & -4.8 & 86.4 & 250 & 0.000 &  0.0 & 96.4 & 250 \\
5 & Gamma & 0.312 & 20.6 &  0.0 & 250 & 0.245 & 19.3 &  2.0 & 250 & 0.093 & 10.2 & 63.6 & 250 & 0.000 &  0.0 & 96.0 & 250 \\
1 & GenF & -0.019 &  2.1 & 94.4 & 248 & -0.010 & -2.1 & 70.8 & 185 & 0.046 & -6.8 & 26.8 & 74 & -0.008 &  0.5 & 85.6 & 232 \\
2 & GenF & -0.057 & -7.7 & 69.6 & 248 & -0.000 &  0.0 & 71.6 & 185 & 0.019 & 96.3 & 38.4 & 98 & 0.000 &  0.0 & 84.4 & 232 \\
3 & GenF & 0.068 &  6.4 & 64.0 & 248 & 0.033 &  3.1 & 63.2 & 185 & -0.012 & -2.3 & 42.4 & 108 & 0.005 & -1.3 & 86.0 & 234 \\
4 & GenF & 0.068 &  5.2 & 67.6 & 248 & 0.038 &  3.2 & 63.2 & 185 & 0.009 &  1.2 & 41.2 & 108 & 0.001 & -2.6 & 87.6 & 237 \\
5 & GenF & 0.001 &  0.1 & 94.8 & 248 & 0.030 &  2.4 & 64.8 & 185 & 0.036 &  3.9 & 40.0 & 108 & -0.012 & -5.2 & 85.6 & 237 \\
1 & FPAFT-df=2 & 0.154 & -16.8 & 35.6 & 250 & -0.166 & -34.6 &  6.8 & 250 & 0.053 & -7.8 & 84.0 & 250 & 0.305 & -17.9 &  7.2 & 250 \\
2 & FPAFT-df=2 & -0.125 & -17.0 & 16.0 & 250 & -0.066 & -7.2 & 67.6 & 250 & 0.010 & 50.7 & 96.8 & 250 & 0.298 & -34.2 &  2.4 & 250 \\
3 & FPAFT-df=2 & 0.053 &  5.0 & 74.4 & 250 & 0.093 &  8.7 & 56.8 & 250 & -0.100 & -18.9 & 44.0 & 250 & 0.292 & -76.0 &  1.2 & 250 \\
4 & FPAFT-df=2 & 0.113 &  8.7 & 54.0 & 250 & 0.206 & 17.4 &  6.4 & 250 & -0.057 & -7.4 & 76.0 & 250 & 0.289 & -740.4 &  1.6 & 250 \\
5 & FPAFT-df=2 & 0.109 &  7.2 & 67.2 & 250 & 0.292 & 23.0 &  0.0 & 250 & 0.027 &  3.0 & 92.0 & 250 & 0.286 & 125.0 &  2.4 & 250 \\
1 & FPAFT-df=3 & 0.090 & -9.8 & 66.8 & 250 & -0.126 & -26.2 & 20.4 & 250 & 0.026 & -3.8 & 94.8 & 250 & 0.308 & -18.1 &  7.6 & 250 \\
2 & FPAFT-df=3 & -0.004 & -0.5 & 92.8 & 250 & -0.108 & -11.8 & 28.0 & 250 & -0.045 & -228.0 & 87.6 & 250 & 0.299 & -34.3 &  5.6 & 250 \\
3 & FPAFT-df=3 & 0.077 &  7.2 & 57.2 & 250 & -0.029 & -2.7 & 84.0 & 250 & -0.113 & -21.4 & 32.0 & 250 & 0.291 & -75.7 &  1.6 & 250 \\
4 & FPAFT-df=3 & 0.015 &  1.2 & 96.0 & 250 & 0.020 &  1.7 & 90.4 & 250 & -0.023 & -3.0 & 91.2 & 250 & 0.287 & -735.3 &  2.0 & 250 \\
5 & FPAFT-df=3 & -0.092 & -6.1 & 72.8 & 250 & 0.054 &  4.2 & 83.6 & 250 & 0.103 & 11.3 & 54.8 & 250 & 0.283 & 123.7 &  4.8 & 250 \\
1 & FPAFT-df=4 & 0.107 & -11.6 & 59.6 & 249 & 0.010 &  2.1 & 88.4 & 250 & -0.050 &  7.4 & 83.6 & 250 & 0.306 & -18.0 &  7.6 & 250 \\
2 & FPAFT-df=4 & 0.005 &  0.7 & 91.6 & 249 & 0.001 &  0.1 & 90.4 & 250 & -0.036 & -182.4 & 89.6 & 250 & 0.303 & -34.8 &  5.2 & 250 \\
3 & FPAFT-df=4 & 0.084 &  7.9 & 52.0 & 249 & 0.017 &  1.6 & 93.6 & 250 & -0.108 & -20.4 & 36.0 & 250 & 0.289 & -75.2 &  2.4 & 250 \\
4 & FPAFT-df=4 & 0.017 &  1.3 & 96.8 & 249 & 0.010 &  0.8 & 94.8 & 250 & -0.055 & -7.1 & 78.0 & 250 & 0.281 & -719.9 &  3.6 & 250 \\
5 & FPAFT-df=4 & -0.095 & -6.3 & 72.0 & 249 & -0.001 & -0.1 & 95.6 & 250 & 0.033 &  3.6 & 91.6 & 250 & 0.274 & 119.8 & 19.6 & 250 \\
1 & FPAFT-df=5 & 0.094 & -10.2 & 66.4 & 250 & 0.016 &  3.3 & 89.6 & 250 & -0.036 &  5.3 & 87.6 & 250 & 0.306 & -18.0 &  8.8 & 250 \\
2 & FPAFT-df=5 & 0.000 &  0.0 & 89.2 & 250 & 0.011 &  1.2 & 92.4 & 250 & -0.025 & -126.7 & 84.8 & 250 & 0.302 & -34.7 &  6.8 & 250 \\
3 & FPAFT-df=5 & 0.080 &  7.5 & 57.2 & 250 & 0.021 &  2.0 & 92.4 & 250 & -0.066 & -12.5 & 65.2 & 250 & 0.289 & -75.2 &  3.6 & 250 \\
4 & FPAFT-df=5 & 0.015 &  1.2 & 96.0 & 250 & 0.009 &  0.8 & 94.8 & 250 & -0.040 & -5.2 & 80.0 & 250 & 0.279 & -714.8 &  5.2 & 250 \\
5 & FPAFT-df=5 & -0.094 & -6.2 & 72.4 & 250 & -0.006 & -0.5 & 96.8 & 250 & 0.013 &  1.4 & 94.4 & 250 & 0.272 & 118.9 & 35.6 & 250 \\
1 & FPAFT-df=6 & 0.088 & -9.6 & 68.8 & 250 & 0.019 &  4.0 & 86.0 & 250 & -0.016 &  2.4 & 93.6 & 250 & 0.306 & -18.0 &  9.2 & 250 \\
2 & FPAFT-df=6 & -0.006 & -0.8 & 86.8 & 250 & 0.019 &  2.1 & 92.0 & 250 & -0.023 & -116.5 & 80.8 & 250 & 0.302 & -34.7 &  8.8 & 250 \\
3 & FPAFT-df=6 & 0.073 &  6.9 & 60.4 & 250 & 0.025 &  2.3 & 92.4 & 250 & -0.040 & -7.6 & 80.8 & 250 & 0.289 & -75.2 &  5.6 & 250 \\
4 & FPAFT-df=6 & 0.015 &  1.2 & 96.8 & 250 & 0.008 &  0.7 & 95.6 & 250 & -0.027 & -3.5 & 87.6 & 250 & 0.279 & -714.8 &  7.6 & 250 \\
5 & FPAFT-df=6 & -0.088 & -5.8 & 76.4 & 250 & -0.011 & -0.9 & 96.4 & 250 & 0.006 &  0.7 & 95.2 & 250 & 0.271 & 118.5 & 38.0 & 250 \\
1 & FPAFT-df=7 & 0.076 & -8.3 & 68.4 & 248 & 0.017 &  3.5 & 86.4 & 250 & -0.011 &  1.6 & 92.0 & 249 & 0.300 & -17.6 & 12.4 & 250 \\
2 & FPAFT-df=7 & -0.012 & -1.6 & 87.6 & 248 & 0.021 &  2.3 & 89.2 & 250 & -0.019 & -96.3 & 83.2 & 249 & 0.296 & -34.0 & 11.2 & 250 \\
3 & FPAFT-df=7 & 0.065 &  6.1 & 68.4 & 248 & 0.027 &  2.5 & 90.0 & 250 & -0.024 & -4.5 & 84.8 & 249 & 0.284 & -73.9 &  8.4 & 250 \\
4 & FPAFT-df=7 & 0.013 &  1.0 & 95.2 & 248 & 0.010 &  0.8 & 96.0 & 250 & -0.018 & -2.3 & 90.8 & 249 & 0.277 & -709.7 & 11.2 & 250 \\
5 & FPAFT-df=7 & -0.084 & -5.6 & 79.2 & 248 & -0.010 & -0.8 & 96.0 & 250 & 0.002 &  0.2 & 93.6 & 249 & 0.271 & 118.5 & 46.0 & 250 \\
1 & FPAFT-df=8 & 0.063 & -6.9 & 72.0 & 249 & 0.014 &  2.9 & 82.8 & 250 & -0.007 &  1.0 & 93.2 & 250 & 0.308 & -18.1 & 15.6 & 249 \\
2 & FPAFT-df=8 & -0.015 & -2.0 & 86.4 & 249 & 0.018 &  2.0 & 88.4 & 250 & -0.013 & -65.9 & 84.8 & 250 & 0.300 & -34.5 & 12.8 & 249 \\
3 & FPAFT-df=8 & 0.057 &  5.4 & 72.8 & 249 & 0.024 &  2.2 & 91.2 & 250 & -0.015 & -2.8 & 89.2 & 250 & 0.288 & -75.0 &  8.8 & 249 \\
4 & FPAFT-df=8 & 0.012 &  0.9 & 94.8 & 249 & 0.008 &  0.7 & 94.0 & 250 & -0.012 & -1.6 & 92.8 & 250 & 0.282 & -722.5 & 15.6 & 249 \\
5 & FPAFT-df=8 & -0.078 & -5.2 & 80.8 & 249 & -0.012 & -0.9 & 94.8 & 250 & 0.001 &  0.1 & 94.8 & 250 & 0.278 & 121.5 & 50.4 & 249 \\
1 & FPAFT-df=9 & 0.057 & -6.2 & 73.6 & 249 & 0.014 &  2.9 & 83.6 & 250 & -0.002 &  0.3 & 91.2 & 249 & 0.299 & -17.6 & 17.2 & 249 \\
2 & FPAFT-df=9 & -0.014 & -1.9 & 86.4 & 249 & 0.016 &  1.8 & 88.8 & 250 & -0.006 & -30.4 & 84.4 & 249 & 0.293 & -33.6 & 10.4 & 249 \\
3 & FPAFT-df=9 & 0.051 &  4.8 & 77.2 & 249 & 0.024 &  2.2 & 90.0 & 250 & -0.008 & -1.5 & 88.0 & 249 & 0.283 & -73.6 &  8.4 & 249 \\
4 & FPAFT-df=9 & 0.012 &  0.9 & 95.6 & 249 & 0.008 &  0.7 & 94.4 & 250 & -0.007 & -0.9 & 92.4 & 249 & 0.281 & -719.9 & 18.4 & 249 \\
5 & FPAFT-df=9 & -0.072 & -4.8 & 85.2 & 249 & -0.010 & -0.8 & 95.6 & 250 & 0.002 &  0.2 & 94.0 & 249 & 0.281 & 122.8 & 54.8 & 249 \\
\hline 
\end{tabular}
}
\end{center}
\end{table}

From Table \ref{tab:sim}, looking at scenarios 1 to 3, the Weibull AFT model gives substantial bias in estimates of the log acceleration factor, and poor coverage probabilities. Similarly, but to a lesser extent, the generalised gamma also indicates some bias and poor coverage, but in addition an important proportion of models, 69 out of 250, failed to converge in Scenario 3 when $\beta=0.5$. The generalised F model performed particularly poorly as a substantial proportional in most scenarios failed to converge; therefore the bias and coverage estimates calculated only on models which converged, should be interpreted with caution. Results based on those that did converge indicate some bias across scenarios, but particularly poor coverage across all scenarios. In all scenarios, the flexible parametric AFT performed well across varying degrees of freedom. In scenarios 1 to 3, there was a FPAFT with a specific degree of freedom (or multiple), that outperformed the Weibull, generalized gamma, and generalized F, both in terms of less bias and coverage probabilities closer to the optimum of 95\%. When there was bias in specific degrees of freedom, the AIC and BIC indicated a more appropriate well fitting model, generally with the least bias. For example, in Scenario 1 with $\beta=-0.5$, the FPAFT with df=2 had -10.6\% bias and coverage probability of 68.4\% and on average was the 10th best fitting model based on both the AIC and BIC; however, all other degrees of freedom were better fitting, for example, with df=7, percentage bias was -0.4\% with coverage of 94.8\%, ranked 2nd and 5th on average in terms of AIC and BIC, respectively. In Scenario 4, where the true model was a Weibull (equivalent to FPAFT with df=1), generally all models estimated the log acceleration factor with minimal bias; however, coverage began to be suboptimum as the degrees of freedom increased in the FPAFT, clearly due to over-fitting. Generally, a flexible parametric AFT model was the best fitting in terms of both AIC and BIC, apart from scenario 4 where the true Weibull model (which is equivalent to a flexible parametric AFT with 1 degree of freedom). In some settings the generalized F was best fitting; however, this is based only on estimates that converged (for example scenario 3 and $\beta=-0.5$, only 106 out of 250 converged).

Moving to estimates of survival in Tables \ref{tab:simpos0} to \ref{tab:simneg1}, in scenarios 1 to 3, the Weibull model produced substantial bias and poor coverage, compared to excellent performance in scenario 4 when the truth was Weibull. Both the generalized gamma and generalized F models produced varying levels of bias and poor coverage, particularly the generalised F, across all 4 scenarios. Both suffered from varying levels of lack of convergence, and also the delta method failed to calculate a standard error in a small number of simulations. The flexible parametric model performed well across all 4 scenarios; there was at least one degree of freedom which provided generally unbiased estimates of survival in each treatment group, with coverage around the 95\% optimum.

\section{Breast cancer in England and Wales}
\label{sec:data}

To illustrate the proposed AFT model, we use a dataset of 9721 women aged under 50 and diagnosed with breast cancer in England and Wales between 1986 and 1990. Our event of interest is death from any cause, where 2,847 events were observed, and we have restricted follow-up to 5 years, leading to 6,850 censored at 5 years. We are interested in the effect of deprivation status, which was categorised into 5 levels; however, in this example we restrict our analyses to comparing the least and most deprived groups. We subsequently have a binary covariate, with 0 for the least deprived and 1 for the most deprived group.

We fit Weibull, generalized gamma, generalized F and the proposed AFT models with 2 to 9 degrees of freedom, and present estimates of the log acceleration factor for the effect of deprivation status, its standard error and associated 95\% confidence interval in Table \ref{tab:df}, and also model fit statistics, namely the AIC and BIC.

\begin{table}[!ht]
\caption{Comparison of parametric AFT models applied to the England and Wales breast cancer dataset.}
\label{tab:df}
\centering
\scalebox{1}{
\begin{tabular}{rrrrrrr}\toprule
 Model & Estimate & Std. Err. & \multicolumn{2}{c}{95\% CI} & AIC & BIC \\
\midrule
Weibull & -0.258 & 0.038 & -0.331 & -0.184 & 17622.17 & 17640.03 \\
Gen. Gamma & -0.287 & 0.041 & -0.367 & -0.207 & 17606.17 & 17629.99 \\
Gen. F & -0.348 & 0.041 & -0.375 & -0.199 & 17544.37 & 17574.14 \\
FPAFT df=2 & -0.263 & 0.039 & -0.339 & -0.188  & 17619.39 & 17643.21 \\
FPAFT df=3 & -0.296 & 0.039 & -0.372 & -0.220  & 17524.85 & 17554.62 \\
FPAFT df=4 & -0.304 & 0.041 & -0.385 & -0.223  & 17526.47 & 17562.20 \\
FPAFT df=5 & -0.307 & 0.042 & -0.390 & -0.224  & 17527.91 & 17569.58 \\
FPAFT df=6 & -0.308 & 0.043 & -0.391 & -0.224  & 17529.84 & 17577.47 \\
FPAFT df=7 & -0.309 & 0.043 & -0.394 & -0.224  & 17531.44 & 17585.03 \\
FPAFT df=8 & -0.323 & 0.042 & -0.405 & -0.242  & 17529.52 & 17589.06 \\
FPAFT df=9 & -0.345 & 0.047 & -0.438 & -0.252  & 17529.18 & 17594.67 \\
\bottomrule 
\end{tabular}
}
\end{table}

Table \ref{tab:df} indicates that the best fitting model, both in terms of lowest AIC and BIC, is the flexible parametric AFT model with 3 degrees of freedom. This estimates an acceleration factor of 0.744 (95\% CI: 0.689, 0.803) for the effect of deprivation status, indicating a patient's survival time is reduced by 25.6\% (95\% CI: 19.7\%, 31.1\%) by being in the most deprived group, compared to the least deprived.

The differences in AIC and BIC are substantial between the best fitting model, and those commonly used, namely the Weibull, gamma and F models. We also observe important variation in the estimates of the effect of deprivation status between the FPAFT with df=3, and the Weibull, gamma and F model estimates.

\begin{figure}[!ht]
	\centering
  \includegraphics[width=0.5\textwidth]{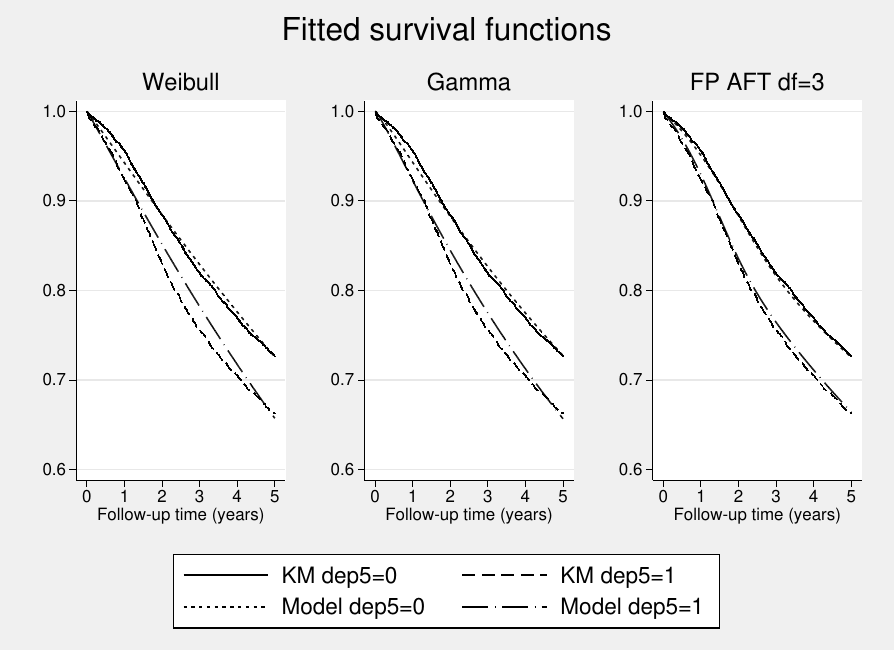}
	\caption{Fitted survival for each deprivation groups, for Weibull, gamma, and Flexible parametric AFT (df=3) models.}
	\label{fig:fitsurv}
\end{figure}

\begin{figure}[!ht]
	\centering
	\begin{subfigure}[b]{0.45\textwidth}
	   \includegraphics[width=\textwidth]{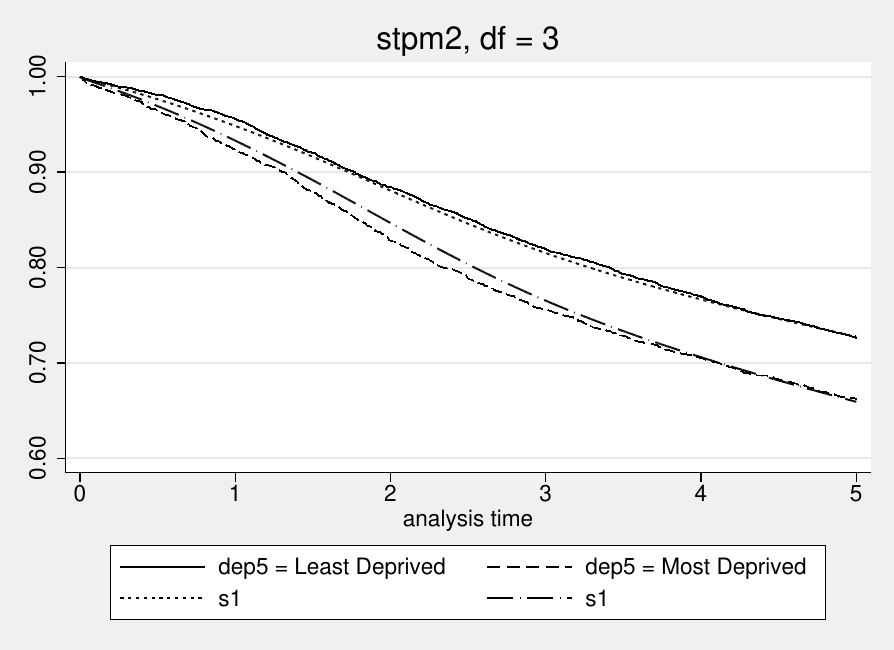}
	   \caption{Proportional hazards model.}
	   \label{fig:stpm2}
	\end{subfigure}
	\begin{subfigure}[b]{0.45\textwidth}
	   \includegraphics[width=\textwidth]{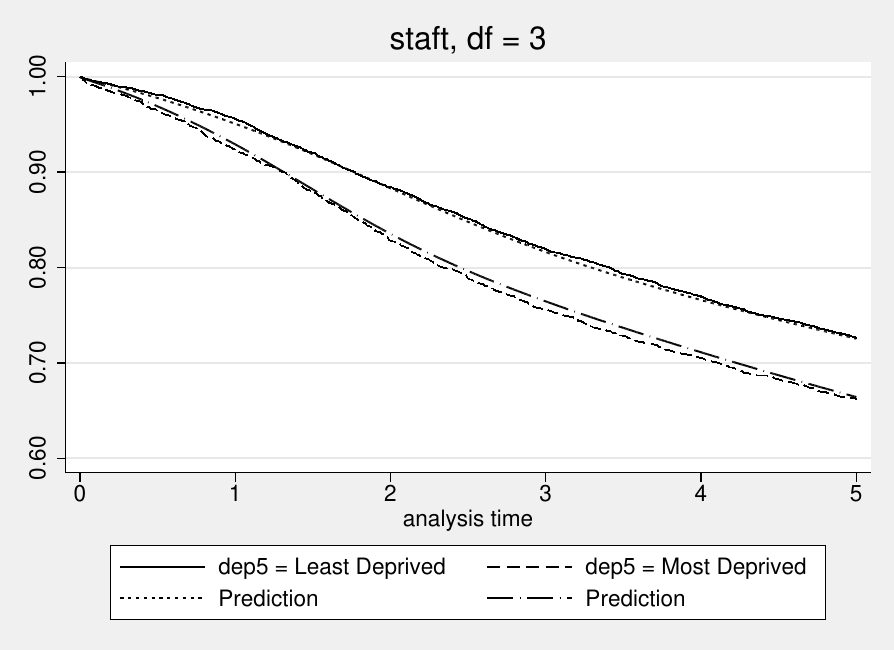}
	   \caption{Accelerated failure time model.}
	   \label{fig:staft}
	\end{subfigure}
	\caption{Fitted survival function for the best fitting flexible parametric models in the proportional hazards and accelerated failure time metrics.}
	\label{fig:fitsurv2}
\end{figure}

We illustrate the fitted AFT models in Figure \ref{fig:fitsurv}, showing the fitted survival function for both deprivation groups, for the Weibull, gamma, F and best fitting flexible AFT model, overlaid on the Kaplan-Meier estimates. It is evident from Figure \ref{fig:fitsurv2} that the flexible AFT fits substantially better than the other models.

\section{Discussion}
\label{sec:disc}

Accelerated failure time models provide an attractive alternative to the proportional hazards framework, particularly for patients, as an acceleration factor can have a more intuitive meaning, directly increasing/decreasing survival time, rather than the event rate. Many authors have argued that AFT models are underused in applied research \citep{Swindell2009,Kay2002,Ng2015}. Indeed, estimates have been shown to be more robust to covariate omission, compared to proportional hazards models \citep{Lambert2004,Keiding1997}. In this article we proposed a new general parametric accelerated failure time model. We focused on the use of restricted cubic splines to provide a highly flexible framework with which to capture complex, biologically plausible functions. Our model can be thought of as an accelerated failure time formulation of that proposed by \cite{Royston2002}. Furthermore, we extended the framework to allow time-dependent acceleration factors. 

Accelerated failure time models show considerable promise for causal inference. In particular, the log acceleration factors are collapsible for omitted covariates that are uncorrelated with the exposure of interest, whereas the proportional hazards models are sensitive to such random effects or frailties. Moreover, the proportional hazards model have a difficult causal interpretation (see Equations~\eqref{eq:causal}). 

We conducted a simulation study to evaluate the performance of the proposed AFT model, indicating excellent performance in a variety of complex, but plausible, settings. In our scenarios, it outperformed the Weibull, generalized gamma and generalized F models, both in terms of minimizing bias in estimates of the acceleration factor and coverage probabilities closer to the optimum 95\%. Furthermore, we found that model selection criteria can aid in selecting degrees of freedom, both to select a model with minimal bias, but also a model which capture the baseline to provide reliable estimates of absolute risk such as survival probabilities. The proposed flexible parametric AFT models is also highly computationally efficient.

An accelerated failure time model will be most appropriate when the covariate effects are multiplicative on a time scale. This scale is intuitive for modelling life expectancy, but is more difficult to interpret in terms of competing risks. To describe this difficulty, consider dividing causes of death into two groups, where an exposure affect the causes of death with different acceleration factors. Then survival from all cause will be the product of two survival probabilities which have ``ageing'' at different rates for the different causes. 

In contrast, the proportional hazards models assume that the covariate effects are multiplicative on the hazards scale. This scale is more intuitive for modelling system dynamics and for competing risks, but these models have a difficult causal interpretation and they are less intuitive for the lay person. As a third model class, the additive hazards model are intuitive for effects operating as competing events and have a straightforward causal interpretation, however they are less intuitive for interpreting effects on the same mechanistic pathway.

Extensions to the framework that would be useful include incorporating random effects, to account for clustered structures and unobserved heterogeneity \citep{Lambert2004,Crowther2014a}, and the extension to interval censoring, possibly in combination with delayed entry.

We provide user-friendly Stata and R software packages to allow researchers to directly use the proposed model framework. For Stata, the command can be installed by typing \texttt{ssc install staft}. For R, the \texttt{rstpm2} package on CRAN provides an \texttt{aft} regression function.

\bibliographystyle{biom}
\bibliography{paper_arxiv.bbl}


\end{document}